\documentstyle[12pt,epsfig]{article}

\begin{document}

\def\ii{\'{\char'20}}
\def\r{\rightarrow}
\def\err{\end{array}}
\def\bea{\begin{eqnarray}}
\def\eea{\end{eqnarray}}
\newcommand{\beq}{\begin{equation}}
\newcommand{\eeq}{\end{equation}}
\newcommand{\nn}{\nonumber}

\begin{titlepage}

\title{\Large \bf Calculation of Wilson loops in 2-dimensional Yang-Mills
theories}

\author{J.M. Aroca$^{a}$\thanks{E-mail address: jomaroca@mat.upc.es}
and Yu. Kubyshin$^{b}$\thanks{On leave of absence from the Institute
for Nuclear Physics,
Moscow State University, 117899 Moscow, Russia.
E-mail address: kubyshin@ualg.pt} \\
\\
$^{a}$ Departament de Matem\`atica Aplicada i Telem\`atica \\
Universitat Polit\`ecnica de Catalunya,
C./ Jordi Girona 1 i 3, Mod. C-3 \\
Campus Nord, 08034 Barcelona, Spain \\
\\
$^{b}$ \'Area Departamental de F{\ii}sica, UCEH, Universidade do Algarve \\
Campus de Gambelas, 8000 Faro, Portugal }

\date{January, 1999}

\maketitle

\begin{abstract}

The vacuum expectation value of the Wilson loop functional in pure
Yang-Mills theory on an arbitrary two-dimensional orientable manifold is
studied. We consider the calculation of this quantity for the abelian
theory in the continuum case and for the arbitrary gauge group and
arbitrary lattice action in the lattice case.
A classification of topological sectors of the theory
and the related classification of the principal fibre bundles
over two-dimensional surfaces are given in terms of a cohomology group.
The contribution of
$SU(2)$-invariant connections to the vacuum expectation value of the
Wilson loop variable is also analyzed and is shown to be similar to the
contribution of monopoles.

\vspace{2mm}

PACS number(s): 11.15.-q

\end{abstract}

\end{titlepage}

\pagestyle{myheadings}
\markright{Calculation of Wilson loops in 2D Y-M theories}
\setcounter{page}{1}

\section{Introduction}

Yang-Mills theory in two dimensions has been object of
intensive studies during almost two decades. It is known that
the classical theory is trivial and the quantum theory has no
propagating degrees of freedom. However, this theory possesses
many interesting properties when formulated on
topologically nontrivial spaces \cite{Witten1}, \cite{Witten2}
and also in the
large $N$ limit ($N$ is related to the rank of the gauge group
$G=SU(N)$) \cite{DK}. A two-dimensional Yang-Mills theory is almost
topological in the sense that, for example, for compact
spacetime manifolds of surface area (two-dimensional volume) $V$
its partition function
depends only on $e^{2} V$, where $e$ is the gauge coupling
constant.
Also, as we will see, the vacuum expectation value of
a Wilson loop depends only on the area of the
region surrounded by the loop and is independent of the points
where the loop is located.
This is a manifestation of the invariance under the
area preserving diffeomorphisms. It is believed that two-dimensional
Yang-Mills theories share some of the important qualitative
features of four-dimensional ones, in particular the area law behaviour.
It was shown recently that in the strong coupling limit they
can be related to lower-dimensional strings \cite{Gross} and
that they play special role in M-theory \cite{YM-M}. All these
have led to the revival of interest in studying two-dimensional
Yang-Mills theories. Rich mathematical structures,
appearing in such theories, were studied in a
number of papers \cite{Witten1}, \cite{Witten2},
\cite{AB-BT1} (see lecture \cite{Moore}
for a review and references). The partition function and the
vacuum expectation values of Wilson loop functionals were
calculated by many authors with various techniques \cite{W-calc} -
\cite{ALMMT}. Physical aspects of these theories, in particular
the analysis of Polyakov loops and the $\theta$-vacua, were
discussed in Refs. \cite{Grig}.

In the present article we focus on the study of the vacuum
expectation values $<T_{\gamma}>$ of the Wilson loop functionals
$T_{\gamma}$ in the pure
Yang-Mills theories with arbitrary gauge group $G$ on
two-dimensional orientable compact manifolds $M$. Such theories
are non-trivial due to non-trivial topology of $M$.
In the classical theory the Wilson loop functionals \cite{Wi}
form a natural (over)complete set of gauge invariant variables.
It is believed
that their vacuum expectation values play similar role
in the quantum theory. The area law behaviour of this quantity
is considered as an indication of the regime of confinement in the
corresponding gauge theory with quarks. There are various techniques
for the calculation of $<T_{\gamma}>$
in two-dimensional pure gauge theories \cite{W-calc} -
\cite{ALMMT}. Here we carry out such calculation by using different
techniques and paying attention to some important details.
In the abelian case $G=U(1)$ we calculate the vacuum expectation value
of the Wilson variable by expanding the functional integral over
the topological sectors of the theory and integrating over
fluctuations around the monopole solution in each sector.
The motivation is based on an analogy with quantum mechanics
which we will also discuss.
It will be shown that the functional integral for $<T_{\gamma}>$ is
indeed saturated by the sum of the contributions of the
topological sectors. For this expansion one needs a classification of
the topological sectors of the theory
or, equivalently, the classification of the principal fibre bundles
over two-dimensional surfaces, and we will give a solution of this
problem in terms of cohomology groups. Next, we perform the
calculation of the vacuum expectation value
of the Wilson variable for an arbitrary gauge group using a lattice
technique. The novelty here is in carrying out the integration
on the lattice in terms of loop variables associated to the plaquettes
and homotopically non-trivial loops and imposing the relation between
the loops on the lattice by inserting a certain
$\delta$-function. An alternative technique of the calculation
in terms of differential forms on the lattice, that works for the
abelian case is also discussed.

Besides discussing the methods of calculation of the
vacuum expectation value $<T_{\gamma}>$ of the Wilson variable
we also focus
on the analysis of the contributions of monopole solutions and
invariant gauge connections \cite{KN}, \cite{Schwarz}
to this quantity. Our aim is to understand the role of such
configurations in the formation of the area law dependence of
$<T_{\gamma}>$. The class of invariant connections plays important
role in certain problems of field
theory, like, for example, in the coset space dimensional reduction
in multidimensional gauge theories \cite{Manton}, \cite{KMRV}.
Besides many known monopole and instanton solutions, it also includes
families of
connections parametrized by a finite number of parameters.
The expectation is that though this class of connections
is rather restricted, it may nevertheless capture some important
properties of the complete theory.
As concrete examples we analyze the invariant connections in the
Yang-Mills theory with certain gauge groups on the two-dimensional
sphere $S^{2}$ and calculate the contribution of the family of
invariant connections in the case of $G=SU(2)$. Our observation
is that the quadratic area dependence of the contribution of the
monopoles and of the invariant connections to $\ln <T_{\gamma}>$
is crucial for the area law dependence of the complete function in
a wide range of the area variable.

The plan of the article is as follows. In Sect. 2 we will give basic
definitions and discuss the functional integral in gauge theory in the case
when the space of connections consists of many components (sectors).
The result on the classification of principle fibre bundles is
also presented. In Sect. 3 we will calculate the vacuum expectation
value $<T_{\gamma}>$ of the Wilson loop variable
in the abelian case for arbitrary $M$. In Sect. 4 $<T_{\gamma}>$
will be calculated on the lattice for an arbitrary gauge group and
arbitrary lattice action. In Sect. 5 we will study the invariant
connections and their contribution to $<T_{\gamma}>$
in the case $M=S^{2}$ and $G=SU(2)$. In Sect. 6 a discussion of the role
of monopoles and invariant connections for the area law
behaviour of $<T_{\gamma}>$ will be presented.

\vspace{0.5cm}

\section{Wilson loop variables and integration over connections}

Let us first set some basic notations and definitions.
In the present article we consider pure gauge theories
on a two-dimensional orientable Riemannian manifold $M$ with
a gauge group $G$. $M$ plays the role of the Euclidean space-time,
and we will mainly be considering the case when it is compact.
By ${\cal G}$ we will denote the Lie algebra of $G$.
We use the formalism of principal fibre bundles and connections
in them (see, for example, \cite{KN}) to describe the gauge theory
\cite{gauge-geom}, \cite{EGH}. Let $P(M,G)$ be a principal
fibre bundle over the manifold $M$ with the
structure group $G$. A potential $A_{\mu}$ of the gauge theory is
characterized by a connection in $P(M,G)$ via a local section in a
standard way. Namely, let $s$ be a local section over a neighbourhood
$U$ of $M$. Then $A = (ie)^{-1} s^{*}w$, where $e$ is the gauge
coupling constant and $s^{*}w$ is the pull-back of the connection form
$w$ with respect to this section (we use for $A$ the normalization
adopted in field theory). The ${\cal G}$-valued
1-form $A$ determines the components of the local gauge potential
$A_{\mu}$ on $M$ in the gauge corresponding to the section $s$ through
the relation $A = A_{\mu} dx^{\mu}$. If $U_{i}$ and $U_{j}$ are
two overlapping neighbourhoods, then at $x \in U_{ij} = U_{i}
\bigcap U_{j}$ the gauge potential forms $A^{(i)}$
and $A^{(j)}$, defined for $U_{i}$ and $U_{j}$ respectively,
satisfy certain consistency conditions. These conditions basically
mean that $A^{(i)}$ and $A^{(i)}$ are gauge equivalent.
{\it Vice versa}, a set of forms $A^{(i)}$, defined
on neighbourhoods $U_{i}$ of an atlas $\{U_{i}, \psi_{i}\}$ of the
manifold $M$ and satisfying the consistency conditions, define a unique
connection form $w$ on $P(M,G)$ (see, for example, \cite{gauge-geom}).
We denote by ${\cal A}$ the space of connections in $P(M,G)$.

Let us fix a point $x_{0}$ in $M$ and consider based loops defined in a
standard way as continuous mappings of the unit interval $I=[0,1]$
into $M$:
\[
\gamma: I \r M, \; \; \; s \in I \r \gamma(s) \in M
\]
with $\gamma(0)=\gamma(1)=x_{0}$. The space of loops on $M$ based
at $x_{0}$ will be denoted as $\Omega (M,x_{0})$.

Let us consider a loop $\gamma$ which, for the sake of simplicity,
is situated inside one neighbourhood and associate with it the element
of $G$ called holonomy:
\beq
H_{\gamma}(A) = {\cal P} \exp \left( ie\oint_{\gamma} A \right),
                 \label{H-def}
\eeq
where $e$ is the gauge coupling constant,
${\cal P}$ means path ordering and $A$ is the gauge 1-form, defining
the potential. We will be interested in traced holonomies
\beq
 T_{\gamma}(A) = \frac{1}{d_{R}} Tr \ R(H_{\gamma}(A)) =
  \frac{1}{d_{R}} \chi_R (H_{\gamma}(A))
   \label{trh}
\eeq
called also Wilson loop variables \cite{Wi}. Here $R$ is an
irreducible representation of $G$, $d_{R}$ is its dimension and
$\chi_R$ is its character. Let us denote by ${\cal T}$ the group
of local gauge transformations, i.e. the group of smooth vertical
automorphisms of $P$. It is known that a set of elements
$T_{\gamma}$, associated to loops $\gamma \in  \Omega (M,x_{0})$ and
satisfying certain relations, called Mandelstam conditions,
suffices to separate points in ${\cal A}/{\cal T}$,
representing classes of gauge equivalent configurations,
and therefore enable to reconstruct all smooth gauge connections
up to gauge equivalence \cite{Giles}. Due to this property the
Wilson loop functionals (\ref{trh}) form a natural set of gauge
invariant functions of connections in the classical theory.

We will study the vacuum expectation
value $<T_{\gamma}>$ of the Wilson loop functional defined as
\bea
< T_{\gamma} > & = & \frac{1}{Z(0)} Z(\gamma), \label{Tvev} \\
  Z(\gamma) & = &  \int {\cal D}A e^{-{\cal S}(A)} T_{\gamma}(A),
 \label{Zg-def}
\eea
where the Yang-Mills action is given by
\beq
{\cal S}(A) = \frac{1}{4}\int_{M} Tr \left( F \wedge \ *F \right) =
\int_{M} \frac{1}{8} Tr \left( F_{\mu \nu} F^{\mu \nu} \right)
\sqrt{ \det g_{\mu \nu}} d^{2}x.
\label{action}
\eeq
The expression (\ref{Zg-def}) and the measure ${\cal D}A$
are understood in the heuristic sense adopted in quantum field
theory. It is known that in two-dimensional Yang-Mills theories
the measure can be defined in a rigorous way \cite{AL} (see also
\cite{ALMMT}), we will return to this issue in Sect. 4.
The integral in (\ref{Zg-def}) is taken over the space of connections
${\cal A}$ (we will not make distinction between connection forms in
$P(M,G)$ and gauge forms $A$ on $M$).
In general ${\cal A}$ may consist of a number of
components (or sectors) ${\cal A}_{\alpha}$, labelled by elements
$\alpha$ of an index set ${\cal B}$,
$\alpha \in {\cal B}$.
The functional integral is represented as a sum
over the elements of ${\cal B}$, each term of the sum
being the functional integral over the connections in ${\cal
A}_{\alpha}$. This feature has an analog in quantum mechanics
which we are going to discuss first.

Consider the functional integral describing propagation
of a particle on a manifold $M$ from a point $x_{0}$ to $x_{1}$
during the time $T$:
\[
  K(T;x_{0},x_{1}) = <x_{1}|e^{-TH}|x_{0}> =
   \frac{1}{Z} \int_{x(0)=x_{0}}^{x(T)=x_{1}} \prod_{t} dx(t)
    \exp \left\{ -\int_{0}^{T} L(x,\dot{x}) dt \right\}.
\]
Here we integrate over all trajectories $x(t)$ (in general, not only over
smooth ones) going from $x_{0}$ to $x_{1}$. Let us focus on the case when
the initial point and the final point coincide:
\beq
  K(T;x_{0},x_{0}) =
   \frac{1}{Z} \int_{x(0)=x_{0}}^{x(T)=x_{0}} {\cal D}x(t)
    \exp \left\{ -\int_{0}^{T} L(x,\dot{x}) dt \right\},
             \label{Kint}
\eeq
so that the integral goes over closed paths in $M$, i.e. over elements
of $\Omega (M,x_{0})$. If $M$ is not simply connected, i.e.
the first homotopy group $\pi_{1}(M,x_{0}) \neq 0 $, then
$\Omega (M,x_{0})$ consists of more than one connected components.
The set of path connected components of  $\Omega (M,x_{0})$ is
in one-to-one correspondence with elements of $\pi_{1}(M,x_{0})$
\cite{Greenberg}. So, the functional integral (\ref{Kint}) becomes
\bea
  K(T;x_{0},x_{0}) & = & \sum_{\alpha \in \pi_{1}(M,x_{0})}
  K^{(\alpha)} (T; x_{0},x_{0})    \label{Kint-exp}  \\
  & = & \sum_{\alpha \in \pi_{1}(M,x_{0})}
   \frac{1}{Z} \int_{x^{(\alpha)}(0)=x_{0}}^{x^{(\alpha)}(T)=x_{0}}
   {\cal D}x^{(\alpha)}(t) \exp \left\{ -\int_{0}^{T}
   L(x^{(\alpha)},\dot{x}^{(\alpha)}) dt \right\},   \nonumber
\eea
and each integral in this sum is taken over paths $x^{(\alpha)}$
belonging to a class corresponding to $\alpha \in \pi_{1}(M,x_{0})$.

As an example let us consider the case $M=S^{1}$ and let
$\varphi$ be the angle parameter of the circle,
$0 \leq \varphi < 2\pi$. The fundamental group $\pi_{1}(S^{1})$
is isomorphic to the abelian group of integer numbers,
$\pi_{1}(S^{1}) \cong \bf{Z}$, therefore
the path-connected components of $\Omega (S^{1},0)$ are labelled by
integers $n$ called winding number. The component with a winding
number $n$ consists of loops which go $n$ times around the
circle $S^{1}$ (counterclockwise for $n > 0$ and clockwise for $n<0$).
Thus, in this case
\[
  K(T;0,0) =  \sum_{n=-\infty}^{\infty}
   \frac{1}{Z} \int_{\varphi^{(n)}(0)=0}^{\varphi^{(n)}(T)=0}
   {\cal D}\varphi^{(n)}(t)
    \exp \left\{ -\int_{0}^{T} L(\varphi^{(n)},\dot{\varphi}^{(n)})
    dt \right\}
\]
(see, for example, Ref. \cite{Asorey83} for a study of the quantum
rotor withing such approach).

An alternative but equivalent way of treating the integral
(\ref{Kint}) is to consider it not over the paths on the manifold
$M$ but on the universal covering space $\tilde{M}$ of $M$
\cite{Schulman}. Such space is simply connected and exists for
every connected manifold $M$. In this case we integrate over
paths in $\tilde{M}$.
We fix an initial point $\tilde{x}^{(0)}_{0}$ which projects
to the point $x_{0} \in M$, $p(\tilde{x}^{(0)}_{0}) = x_{0}$, where
$p$ is the covering space map.  Then we sum contributions over
all paths which end at points $\tilde{x}^{(\alpha)}_{0}$
of the set $p^{-1}(x_{0})$, since they all project
to $x_{0}$. It is known that elements of $p^{-1}(x_{0})$ are in
1-1 correspondence with elements $\alpha\in\pi_{1}(M,x_{0})$,
therefore
\[
  K(T;x_{0},x_{0}) =  \sum_{\alpha \in \pi_{1}(M,x_{0})}
   \frac{1}{Z} \int_{\tilde{x}(0)=\tilde{x}^{(0)}_{0}}^{\tilde{x}(T)
  =\tilde{x}^{(\alpha)}_{0}} {\cal D}\tilde{x}(t)
    \exp \left\{ -\int_{0}^{T} L(\tilde{x},\dot{\tilde{x}}) dt \right\}.
\]
In the case $M=S^{1}$ the universal covering space
$\tilde{M}=\bf{R}^{1}$ with the coordinate $\tilde{\varphi}$,
$-\infty < \tilde{\varphi} < \infty$, $p(\tilde{\varphi}) =
\tilde{\varphi} (\mbox{mod} \; 2\pi)$ and
\[
 K(T;0,0) =  \sum_{n=-\infty}^{\infty} \frac{1}{Z}
 \int_{\tilde{\varphi}(0)=0}^{\tilde{\varphi}(T)=2\pi n}
 {\cal D}\tilde{\varphi}(t)
 \exp \left\{ -\int_{0}^{T} L(\tilde{\varphi}, \dot{\tilde{\varphi}})
 dt \right\}.
\]
In principle, to each partial amplitude some weights can be
assigned:
\[
  K(T;x_{0},x_{0}) = \sum_{\alpha \in \pi_{1}(M,x_{0})} \chi (\alpha)
     K^{(\alpha)}(T;x_{0},x_{0})
\]
(cf. (\ref{Kint-exp})). In Ref. \cite{Laidlaw} an analysis of
these weights was carried out and it was shown that this arbitrariness
leads only to an irrelevant overall phase factor for the total
amplitude.

Let us return to the gauge theory. In this case the set of connected
components of ${\cal A}$ is in 1-1 correspondence with the set of
non-equivalent (i.e. which cannot be mapped one into another
by a bundle isomorphism) principal $G$-bundles $P(M,G)$ over
manifold $M$. Let us denote this set as ${\cal B}_{G}(M)$,
i.e. ${\cal B} \cong {\cal B}_{G}(M)$. The problem of
characterization of this set and classification of such bundles is
considered in a number of books and articles. A method,
which in many cases gives a solution to this problem and which we
closely follow here, is discussed in lectures \cite{AI}.
Relevant results from algebraic topology can be found in
\cite{Span}-\cite{Bott}. Here we only outline the construction.
We would like to note that in fact the considerations and the results
in the rest of this section are valid for a more general case,
namely when $M$ is a CW-complex and not just a smooth manifold.

It is known that for each structure group there exists a universal
principal $G$-bundle $EG=P(BG,G)$ with the base $BG$. The total
space $EG$ of the universal bundle
is $\infty$-universal, i.e. $\pi_{q}(EG)=0$ for all $q \geq 1$.
The important property
is that every principal $G$-bundle over $M$ can be obtained (up to
equivalence) from $EG$ as the induced bundle, i.e. $P = f^{*}(EG)$
for some mapping $f: M \rightarrow BG$ \cite{Swit}. One can prove
that ${\cal B}_{G}(M)$ is isomorphic to the space of all homotopy
equivalent classes of mappings $f: M \rightarrow BG$,
\[
    {\cal B}_{G}(M) \cong [M \ ;\ BG].
\]
By examing a certain exact homotopy sequence it can be shown
that the base $BG$ of the universal bundle satisfies the
property \cite{AI}, \cite{Swit}
\beq
       \pi_{q}(BG) = \pi_{q-1}(G), \; \; \; q \geq 1.   \label{piBG}
\eeq
The question is how one can construct such universal bundles
and characterize $[M;BG]$ in terms of objects which can be calculated
in a relatively easy way. It turns out that the Eilenberg - MacLane
spaces play important role for this problem because of their special
homotopic properties. Such spaces are often denoted as $K(\pi,n)$,
where $\pi$ is a group and $n$ is a positive integer, and are
defined as follows:
\bea
 \mbox{i)} & & \; \; K(\pi,n) \; \; \; \mbox{is path connected};
 \nonumber \\
 \mbox{ii)} & & \pi_{q}\left( K(\pi,n) \right) = \left\{
      \begin{array}{ll}
                 \pi, & \mbox{if} \; \;  q=n, \\
                  0 , & \mbox{if} \; \;  q \neq n.
      \end{array} \right.   \nonumber
\eea
If $n \geq 1$ and $\pi$ is Abelian, the space $K(\pi,n)$
exists as a CW-complex and
can be constructed uniquely up to a homotopy equivalence \cite{Span}.
The property, which is crucially important for us, is the following:
\[
   [M;K(\pi,n)] \cong H^{n}(M,\pi),
\]
where $H^{n}(M,\pi)$ is the $n$th singular cohomology group \cite{Span},
\cite{Swit}, \cite{Bredon}. A simple example is $K(Z,1) = S^{1}$.

Often the Eilenberg-Maclane spaces are infinite dimensional.
The example important for us is the space $CP_{\infty}$ which
is a CW-complex. It is understood as the union (direct limit)
of the complex projective
spaces $CP_{n}$ of the sequence $CP_{1} \subset CP_{2} \subset \ldots $
\cite{Span}. Then $\pi_{q} (CP_{\infty}) = \lim _{j \rightarrow \infty}
\pi_{q} (CP_{j})$ and
\[
\pi_{2}(CP_{\infty}) = Z, \; \; \; \pi_{q}(CP_{\infty}) = 0 \; \;
\mbox{for} \; \; q \neq 2.
\]
Thus, $CP_{\infty} = K(Z,2)$.

This space serves for the classification of principal fiber bundles
with $G=U(1)$. Indeed, one can construct the sequence of Hopf fibrations
$S^{2n+1} \rightarrow CP_{n}$ with the fibre $U(1)$. It can be shown that
the space $S^{\infty}$ is $\infty$-universal and the bundle
$S^{\infty} = P(CP_{\infty},U(1))$ is the universal bundle.
Therefore, $EU(1) = S^{\infty}$, $BU(1) = CP_{\infty}$
and
\beq
{\cal B}_{U(1)} (M) \cong [M;BU(1)] = [M;K(Z,2)] \cong H^{2}(M;Z),
\label{BU1}
\eeq
Ref. \cite{AI}.
This relation gives the classification of principal $U(1)$-bundles
$P(M,U(1))$ over $M$. Note that the result (\ref{BU1}) is valid for $M$
of any dimension. For the case when $M$ is a smooth manifold it
was discussed in \cite{Kost-AsBo}.

Now let us consider the generalization of this construction
for the case of other gauge groups. The idea is that for classification
of bundles with the base $M$ with $\dim M \leq n$ only homotopy
groups of low dimensions are important. Then instead of $BG$
one can use some other space $BG_{n}$ which is related to it and
which may be easier to construct and to study.
To define $BG_{n}$ let us introduce the notion of $p$-equivalence.
Consider two path connected spaces $X$ and $Y$ and a continuous
map $f: X \rightarrow Y$ such that for all $x \in X$ the induced map
$f_{*}: \pi_{q} (X,x) \rightarrow \pi_{q}(Y,f(x))$
is an isomorphism for $0<q<p$
and an epimorphism for $q=p$. Such map $f$ is called $p$-equivalence.
Then, if there exists some space $BG_{n}$ and a map $f : BG \rightarrow
BG_{n}$ which is $(n+1)$-equivalence, it can be proved that
\[
   [M;BG] \cong [M; BG_{n}]
\]
for any CW-complex $M$ with $\dim M \leq n$ \cite{Swit}, \cite{Bott}.

Let us suppose that the structure group (gauge group) is connected,
so that $\pi_{0}(G)=0$. Due to Eq. (\ref{piBG}) $\pi_{1}(BG)=0$.
Then we take $BG_{2} = K(\pi_{1}(G),2)$. We have
\bea
\pi_{1}(BG_{2}) & = & \pi_{1}\left(K(\pi_{1}(G),2)\right) = 0
\cong \pi_{1}(BG), \nonumber \\
\pi_{2}(BG_{2}) & = & \pi_{2}\left(K(\pi_{1}(G),2)\right) =
\pi_{1}(G) \cong \pi_{2}(BG),
                                          \nonumber \\
\pi_{q}(BG_{2}) & = & \pi_{q}\left(K(\pi_{1}(G),2)\right) = 0 \; \;
   \mbox{for} \; \; \; q > 2.   \nonumber
\eea
Moreover,
\[
 [BG;BG_{2}] = [BG; K(\pi_{1}(G),2)] \cong H^{2}(BG, \pi_{1}(G))
\]
and this assures the existence of a mapping
$f: BG \rightarrow K(\pi_{1}(G),2)$ which is 3-equivalence
\cite{Span}. Thus, if $\dim M = 2$
\beq
   {\cal B}_{G}(M) \cong [M;BG] \cong [M;BG_{2}] = [M; K(\pi_{1}(G),2)]
   \cong H^{2} (M,\pi_{1}(G)).    \label{BG}
\eeq
This gives us the classification of principal fibre bundles $P(M,G)$
over two-dimensional manifolds CW-complexes
with the structure group $G$.

An equivalent classification of principal fibre bundles over a
two-dimensional surface in terms of elements of the group
$\Gamma$, specifying the global structure of $G$ through the
relation $G = \tilde{G}/\Gamma$, where $\tilde{G}$ is the
universal covering group of $G$, was given in
Ref. \cite{Witten2}.

For completeness, we present here a list of the first homotopy groups
$\pi_{1}(G)$ for  some Lie groups which are of interest in
gauge theories:

\begin{enumerate}

 \item Simply connected, $G = SU(n)$, $Sp(n)$: $\pi_{1}(G) = 0$.

 \item $G = SO(n)$, $n = 3$ and $n \geq 5$: $\pi_{1}(G) = Z_{2}$.

 \item $G=U(n)$: $\pi_{1}(G) = Z$.

\end{enumerate}
For $G=U(n)$ and $n=1$ the gauge group is abelian and
result (\ref{BG}) becomes the relation (\ref{BU1}) discussed above.

\section{Calculation of the Wilson loop for $G=U(1)$, continuous case}

In this section we calculate the vacuum expectation value (\ref{Tvev})
of the Wilson loop variable in the abelian case. Since the surface
$M$ has non-trivial topology the result is non trivial. The functional
integral will be evaluated by summing over the monopoles and
integrating over the
fluctuations around each monopole. The result will be shown to be
exact by comparing with the result of the calculation of $<T_{\gamma}>$
on the lattice in Sect. 4. This means that
the functional integral (\ref{Zg-def}) is saturated by quasiclassical
contributions.

As it was shown in the previous section, $U(1)$-bundles are
classified by elements of the second cohomology group $H^{2}(M,Z)$.
The class $c_{1}(P) \in H^{2}(M,Z)$, corresponding to the bundle $P$,
is known as the first (integer) Chern class \cite{KN},
\cite{Span}, \cite{Bott}. For a closed orientable
2-dimensional manifold $M$ $H^{2}(M,Z) \cong Z$.
For the bundle $P_{n}$, characterized by the label $n \in {\cal B} \cong
Z$, the integer cohomology class can be represented by
$e F^{(n)} / (2 \pi)$ with $F^{(n)}$ being the curvature 2-form
defined locally through $F^{(n)} = dA^{(n)}$, where $A^{(n)}$ is the
gauge 1-form given by a connection in $P_{n}$. The integral
\beq
    \frac{e}{2 \pi} \int_{M}  F^{(n)} = n,     \label{int-c}
\eeq
does not depend on the choice of the connection and gives the
first Chern number which is also called topological charge.

Let us turn to the calculation of the functional integral (\ref{Zg-def}).
We will consider the case of one simple loop
$\gamma$ which is the boundary of a region $\sigma$, i.e.
$\gamma = \partial \sigma$. So, we consider the case of homologically
trivial loop. Generalization to the case of multiple loops
is straightforward though rather cumbersome, see Ref. \cite{ALMMT}.
Let us assume that the Wilson loop variable (\ref{trh}) is defined for the
irreducible respresentation of $U(1)$ labelled by $\nu \in Z$, i.e.
\beq
T_{\gamma}(A) = \exp\left\{ i\nu e \oint_{\gamma} A \right\}. \label{Tnu}
\eeq
Using the quantum mechanics analogy, discussed in the previous
section, and the fact that $H^{2}(M;Z) \cong Z$ we write
\beq
Z(\gamma) = \sum_{n=-\infty}^{\infty} \int {\cal D} A^{(n)}
   e^{-{\cal S}(A^{(n)})} T_{\gamma}(A^{(n)}) = \sum_{n=-\infty}^{\infty}
    Z^{(n)}(\gamma),          \label{Z-sum}
\eeq
where in each term $Z^{(n)}$ the integration goes over gauge
potentials with the Chern number equal to $n$, i.e. over potentials
given by connections in the fibre bundle $P_{n}$. We represent
$A^{(n)} = \tilde{A}^{(n)} + a$, where $\tilde{A}^{(n)}$ is a potential
with
\[
     \int_{M} \tilde{F}^{(n)} = \frac{2\pi n}{e},
\]
$\tilde{F}^{(n)} = d \tilde{A}^{(n)}$ and $a$ is a 1-form with
the zero Chern number. Thus the 2-form $F^{(n)}$ is equal to
$F^{(n)} = \tilde{F}^{(n)} + f$,
where $f = da$ globally, and satisfies  Eq. (\ref{int-c}).
As $\tilde{A}^{(n)}$ we take instanton
configurations, i.e. solutions of the equation of motion. In the
literature they are often referred to as monopoles, and we will
follow this terminology in the present paper.

Assume that the manifold $M$ is endowed with a Riemannian metric
$g_{ij}$. Let us denote by $(\alpha, \beta)$ the scalar
product of forms defined in the standard way:
\beq
(\alpha, \beta) = \int_{M} \alpha \wedge * \beta   \label{inner}
\eeq
for two $p$-forms $\alpha$ and $\beta$.

For the action (\ref{action}) in the abelian case
the equation of motion is $\delta F = 0$, where $\delta$ is the
adjoint of the exterior derivative defined by
$(\alpha, d \beta) = (\delta \alpha, \beta)$. Since any 2-form on the
two-dimensional manifold $M$ is proportional to the volume
2-form $\mu$, which locally is equal to
\beq
   \mu := \sqrt{\det g_{\mu \nu}} dx^{1} \wedge dx^{2},     \label{dv}
\eeq
the 2-form $\tilde{F}^{(n)}$ can be written as
\beq
   \tilde{F}^{(n)} = \kappa^{(n)}(x) \mu. \label{F-kappa}
 \eeq
The equation of motion requires that
\[
  \delta \tilde{F}^{(n)} = -*d* (\kappa^{(n)}(x) dv)
  = * d \kappa^{(n)} = 0,
\]
i.e. $\kappa^{(n)}$ is constant. Then condition (\ref{int-c}) gives
\beq
\kappa^{(n)} = \frac{2\pi n}{Ve},   \label{kappa}
\eeq
where $V$ is the volume (area) of $M$.

The term $Z^{(n)}$ in Eq. (\ref{Z-sum}) equals
\[
   Z^{(n)}(\gamma) = \int_{{\cal A}^{(0)}} {\cal D}a \;
   e^{-{\cal S}(\tilde{A}^{(n)} + a)} \;
   T_{\gamma}(\tilde{A}^{(n)} + a).
\]
Here the functional integration goes over connections
with zero topological charge on the trivial bundle $P_{0}$.
The abelian action can be written as
\bea
{\cal S}(\tilde{A}^{(n)} + a) & = & \frac{1}{2}(F,F) =
\frac{1}{2}(\tilde{F}^{(n)} + f,\tilde{F}^{(n)} + f) \nonumber \\
  & = & \frac{1}{2}(\tilde{F}^{(n)},\tilde{F}^{(n)} ) +
  (\tilde{F}^{(n)},f) + \frac{1}{2}(f,f) \nonumber \\
  & = & {\cal S}(\tilde{A}^{(n)}) + {\cal S}(a)     \nonumber
\eea
since the term linear in $f$ vanishes (recall that $f=da$ globally).
In accordance with definition (\ref{Tnu})
\[
 T_{\gamma}(\tilde{A}^{(n)} + a) = \exp\left\{ i\nu e \oint_{\gamma}
(\tilde{A}^{(n)} + a) \right\}
 = \exp \left\{ i\nu e \oint_{\gamma}\tilde{A}^{(n)}\right\}
\exp \left\{ i\nu e \oint_{\gamma} a \right\},
\]
and the expression for the vacuum expectation value of the Wilson
loop variable factorizes:
\bea
Z(\gamma) & = & \left[ \sum_{n=-\infty}^{\infty} Z_{mon}^{(n)}(\gamma)
\right] Z_{0}(\gamma) ,  \label{Z-fact} \\
Z_{mon}^{(n)} & = & e^{-{\cal S}(\tilde{A}^{(n)})} T_{\gamma}
(\tilde{A}^{(n)}),
     \label{Zn}  \\
Z_{0}(\gamma) & = & \int_{{\cal A}^{(0)}} {\cal D}a \: e^{-{\cal S}(a)}
   T_{\gamma}(a).   \label{Z0}
\eea

Let us calculate first the monopole contribution (\ref{Zn}).
It turns out that for this we
do not need explicit expressions for the monopole solutions
$\tilde{A}^{(n)}$.
Indeed, using Eqs. (\ref{F-kappa}) and (\ref{kappa})
one readily obtains that
\bea
{\cal S}(\tilde{A}^{(n)}) & = &
\frac{1}{2} \int_{M} \tilde{F}^{(n)} \wedge * \tilde{F}^{(n)}
= \frac{1}{2}\int_{M} \kappa^{(n)} \mu \wedge *\kappa^{(n)} \mu
\nonumber \\
 & = & \frac{1}{2}(\kappa^{(n)} )^{2} \int_{M} \mu =
 \frac{1}{2}\left( \frac{2\pi n}{eV} \right)^{2} V
 = \frac{2\pi^{2} n^{2}}{e^{2}V}.       \nonumber
\eea
If $\gamma = \partial \sigma$, where $\sigma$ is the interior
of $\gamma$, and $\tilde{A}^{(n)}$ is regular in $\sigma$,
then by using Stoke's theorem we obtain that
\bea
\oint_{\gamma} \tilde{A}^{(n)} & = & \oint_{\partial \sigma}
\tilde{A}^{(n)} = \int_{\sigma} d\tilde{A}^{(n)}     \nonumber \\
& = &  \int_{\sigma} \tilde{F}^{(n)} = \kappa^{(n)} \int_{S} \mu =
\frac{2\pi n}{eV} S,
\nonumber
\eea
where we denoted the area of the region $\sigma$ by $S$.
We will see that in fact for closed
$M$ the final answer does not depend on whether
we take the interior of
the closed path $\gamma$ (with the area $S$) or its exterior
(with the area $V-S$).
The contribution of the monopoles to the function (\ref{Z-fact})
is equal to
\beq
 Z_{mon}(\gamma) := \sum_{n=-\infty}^{\infty} Z_{mon}^{(n)}(\gamma) =
  \sum_{n=-\infty}^{\infty} \exp \left( -\frac{2\pi^{2}n^{2}}{e^{2}V} +
  i \frac{2\pi n}{V}\nu S \right).   \label{Zmon}
\eeq

For the purpose of comparison with the result of the lattice
calculation and in order to have an expression with clearer symmetry
properties we present Eq. (\ref{Zmon}) in another form by using the
Poisson summation formula
\beq
  \sum_{n=-\infty}^{\infty} f(n) = \sum_{l=-\infty}^{\infty}
\int_{-\infty}^{\infty}
   dz f(z) e^{2\pi i l z}   \label{Poisson}
\eeq
with
\[
f(z) = \exp \left(-\frac{2\pi^{2} z^{2}}{e^{2}V} +
 i \frac{2\pi z}{V}\nu S \right) .
\]
By a straightforward calculation we obtain the following result
\bea
Z_{mon}(\gamma) & = & \sum_{l=-\infty}^{\infty} \int_{-\infty}^{\infty}
dz \exp \left(-\frac{2\pi^{2} z^{2}}{e^{2}V} + 2\pi i
\frac{S}{V}\nu z + 2\pi i l z \right)
  \nonumber \\
  & = & e \sqrt{\frac{V}{2\pi}} \sum_{l=-\infty}^{\infty}
  \exp \left[ - \frac{e^{2}V}{2} \left( \frac{S}{V}\nu +
l \right)^{2} \right].  \label{Zmon1}
\eea
We see that $Z_{mon}$ is a function of $e^{2}V$ and $S/V$. Now it
is clear that it
is invariant under the transformation $S \rightarrow (V-S)$.

As an illustration of this general calculation let us consider a
concrete example of $M=S^{2}$ with $\tilde{A}^{(n)}$ being the
Dirac monopoles. Consider a standard description of the
principal $U(1)$-bundles $P(S^{2},U(1))$ (see, for example,
\cite{EGH}, \cite{DNF}). We cover the sphere $S^{2}$, parametrized
by angles $\vartheta$ and $\varphi$ ($0 \leq \vartheta \leq
\pi$, $0 \leq \varphi < 2\pi$), with two charts, $U_{1}$ and $U_{2}$,
chosen as $U_{1} = S^{2}-\{S\}$ (the sphere without the southern pole)
and $U_{1} = S^{2}-\{N\}$ (the sphere without the northern pole):
\bea
U_{1} & = & \{ 0 \leq \vartheta < \pi, \; 0 \leq \varphi
< 2\pi \}, \label{chart1} \\
U_{2} & = & \{ 0 < \vartheta \leq \pi, \; 0 \leq
\varphi < 2\pi \}.   \label{chart2}
\eea
The fiber $F=U(1) \cong S^{1}$ is parametrized by $e^{i\psi}$.
Thus
\bea
\pi^{-1}(U_{1}) & = & U_{1} \times U(1) = \{\vartheta,\varphi,
\lambda_{+}(\vartheta,\varphi)=e^{i\psi_{1}} \}, \nonumber \\
\pi^{-1}(U_{2}) & = & U_{2} \times U(1) = \{\vartheta,\varphi,
\lambda_{-}(\vartheta,\varphi)= e^{i\psi_{2}} \}, \nonumber
\eea
where $\pi$ is the canonical projection.
The bundle is characterized by the transition function
$\lambda_{12} (\vartheta,\varphi) = \lambda_{2}(\vartheta,
\varphi) \lambda_{1}^{-1}(\vartheta, \varphi)$ which is a
mapping from $U_{12} = U_{1} \cap U_{2}$ to the gauge group $G=U(1)$.
In our case
it is enough to consider the transition function only on the equator,
i.e. the mapping $\tilde{\lambda}_{12}: S^{1} \rightarrow
U(1) \cong S^{1}$. Such mappings are labelled by integers
$n \in Z$ and are of the form
$\tilde{\lambda}_{12}^{(n)}(\varphi) = e^{in\varphi}$. This implies
that the principal bundles $P(S^{2},U(1))$ are labelled by integers
$n$ (in accordance with the general result (\ref{BU1})), and
$\psi_{2} = \psi_{1} + n \varphi$.

Connections in the bundles are given by
\bea
\omega & = & - \frac{i}{2\pi} d\psi_{2} + ie A_{\mu}^{(1)} dx^{\mu}
\; \; \; \mbox{in}
 \; \; \; \pi^{-1}(U_{1}), \nonumber \\
\omega & = & - \frac{i}{2\pi} d\psi_{2} + ie A_{\mu}^{(2)} dx^{\mu}
\; \; \; \mbox{in}
 \; \; \; \pi^{-1}(U_{2}), \nonumber
\eea
and $A_{\vartheta}^{1} = A_{\vartheta}^{2}$, $A_{\varphi}^{1} =
A_{\varphi}^{2} - n/(2\pi)$.
The monopole solution is described by
\beq
  \tilde{A}^{(1)}_{n} = \frac{n}{2e} (1 - \cos \vartheta) d \varphi,
  \; \;
  \tilde{A}^{(2)}_{n} = - \frac{n}{2e} (1 + \cos \vartheta) d \varphi
  \label{U1-mon}
\eeq
and
\beq
  \tilde{F}_{n} = d \tilde{A}^{(1)}_{n} = d \tilde{A}^{(2)}_{n}
   =\frac{n}{2e} \sin
\vartheta d\vartheta \wedge
  d \varphi = \frac{2\pi n}{Ve} \mu   \label{F-U1-mon}
\eeq
in accordance with Eq. (\ref{F-kappa}).
In Eqs. (\ref{U1-mon}), (\ref{F-U1-mon}) the lower index $n$ of
the forms $\tilde{A}^{(i)}_{n}$ and $\tilde{F}_{n}$ is the label of the
fibre bundle in which the corresponding connection is defined.

Now let us turn to the calculation of the contribution
$Z_{0}(\gamma)$ given by Eq. (\ref{Z0}). The action
\[
{\cal S}(a) =  \frac{1}{2} \int_{M} f \wedge *f = \frac{1}{2}(f,f) =
\frac{1}{2} (da, da)
   = \frac{1}{2}(a, \delta d a).
\]
For further discussion we need to consider de Rham currents
\cite{dR-Gelf} or weak forms, according to the
terminology of Ref. \cite{Fels}. Let us denote by $W^{p}(M)$ the
space of weak $p$-forms on $M$ and remind the definition.
Generalizing the notion of distribution a weak $p$-form $K$ is a
linear functional on the space $D^{p}(M)$, i.e. $K: \; D^{p}(M)
\rightarrow R$, where $D^{p}(M)$ is the space of test $p$-forms
with compact support. The space $W^{p}(M)$ includes functionals
corresponding to $C^{\infty}$
$p$-forms, and we will denote the space of such forms
by $\Lambda ^{p}(M)$. If $w \in \Lambda ^{p}(M)$,
then the corresponding functional is defined as
\[
    w[u] := (w,u) \equiv \int_{M} w \wedge * u,
\]
$u \in D^{p}(M)$. Because of this often the
notation $(K,u)$ for $K[u]$ is used, where $K \in W^{p}(M)$.
Some basic operations on forms are extended to weak forms in an
obvious way \cite{Fels}
\beq
d K [u] = K[\delta u], \; \; \delta K [u] = K[d u], \; \;
\Delta K [u] = K[\Delta u],  \label{weak1}
\eeq
etc., where $\Delta = d \delta + \delta d$ is the Laplacian.
If $M$ is compact, which is the case under consideration in this
section, $D^{p}(M) = \Lambda^{p}(M)$.

For our calculation we introduce the weak 1-form $J_{\gamma}$,
characterizing the loop $\gamma$ in the following way:
\[
J_{\gamma}[u] = \oint_{\gamma} u,
\]
$u \in \Lambda^{1}(M)$. Since $\oint_{\gamma} u = \oint_{\gamma}
u_{\nu} dx^{\nu}$, formally we can write
\[
J_{\gamma}[u] = \int_{M} J_{\gamma} \wedge * u =
\int_{M} u_{\nu}(x) J_{\gamma}^{\nu}(x) \: \mu,
\]
where $\mu$ is the volume 2-form (\ref{dv}) and the contravariant
components $J_{\gamma}^{\nu}$ are given by
\beq
  J_{\gamma}^{\nu}(y) = \oint_{\gamma} dx^{\nu} \frac{1}
{\sqrt{\det g_{\mu \nu}}} \delta (y-x).    \label{Jg-delta}
\eeq

We also introduce the weak 2-form $J_{\sigma}$
which characterizes the surface
$\sigma$, the interior of the closed path $\gamma$, as follows:
\beq
J_{\sigma}[v] = \int_{\sigma} v,   \label{JS-def}
\eeq
$v \in \Lambda^{2}(M)$. Since $\gamma = \partial \sigma$, for $u \in
\Lambda^{1}(M)$ we have:
\[
J_{\sigma}[du] = \int_{\sigma} du = \oint_{\partial \sigma} u =
\oint_{\gamma} u = J_{\gamma}[u].
\]
{}From this relation and Eqs. (\ref{weak1}) we conclude
that $\delta J_{\sigma}[u] = J_{\sigma}[du] = J_{\gamma}[u]$, i.e.
\beq
  J_{\gamma} = \delta J_{\sigma}.  \label{Jg-JS}
\eeq
It is easy to see that formally
\beq
   J_{\sigma} = h_{\sigma}(x) \mu ,   \label{JS1}
\eeq
where $h_{\sigma}(x)$ is the characteristic function of the
region $\sigma$ given by
\beq
  h_{\sigma}(x) = \left\{
      \begin{array}{ll}
                 1, & \; \; x \in \sigma, \\
                 0, & \; \;  x \not\in \sigma.
      \end{array} \right.   \label{JS2}
\eeq
Note that in fact $J_{\sigma}$ belongs to
the space ${\cal H}^{2}(M)$ of quadratically integrable 2-forms, i.e.
2-forms for which the norm $ ||J_{\sigma}||$, defined by (\ref{inner}),
is finite.

The contribution of fluctuations can be written as
\beq
Z_{0}(\gamma) = \int_{{\cal A}^{(0)}} {\cal D}a \exp \left[ -
\frac{1}{2} (a, \delta d a)
+ i \nu e (J_{\gamma},a) \right].   \label{Z01}
\eeq
The operator $\delta d$ is not invertible on the space of gauge
1-forms $a$.
First of all there are zero modes due to the gauge invariance
of the theory, that is the symmetry
under gauge transformations $a \rightarrow a'= a + d \xi$.
One deals with this problem
in the standard way by inserting the Faddeev-Popov
$\delta$-function, which gives
rise to the gauge-fixing term, and integrating over the gauge
group \cite{FS}.
Here we consider the covariant gauges, so that the
additional factor equal to $\exp \left[
-(1/2\alpha) (\delta a, \delta a) \right]$, where
$\alpha$ is the gauge fixing parameter, appears in Eq. (\ref{Z01}).
With this addition the quadratic operator does not have gauge zero
modes anymore. We choose $\alpha = 1$, the Feynman gauge. As a
result Eq.(\ref{Z01}) becomes
\beq
Z_{0}(\gamma) = \int_{{\cal A}^{(0)}} {\cal D}a \exp \left[ -
\frac{1}{2} (a, \Delta a)
+ i\nu e (J_{\gamma},a) \right],    \label{Z02}
\eeq
where $\Delta $ is the Laplacian.
Let us remind that since we are considering the abelian case
the Faddeev-Popov determinant gives an irrelevant constant factor
in (\ref{Z02}).

The Laplacian $\Delta$ may have further zero modes
related to symmetries of the
manifold $M$. The zero modes $\alpha_{0}$ are harmonic
1-forms on $M$, i.e. they satisfy $\Delta \alpha_{0} = 0$. We will
denote the space of harmonic $p$-forms on $M$ as $Harm^{p}(M)$.
For example, if $M$ is the
two-dimensional torus $T^{2}$, parametrized by the angles
$(\varphi_{1}, \varphi_{2})$
($0 \leq \varphi^{i} < 2\pi$), there are two independent zero modes
$\alpha_{01}= d\varphi_{1}/(2\pi)$ and $\alpha_{02}=
d\varphi_{2}/(2\pi)$. Any 1-form
$\alpha = c_{1} \alpha_{01} + c_{2} \alpha_{02}$, where
$c_{1}$ and $c_{2}$ are constants, is harmonic.
It cannot be eliminated by
gauge transformation $\alpha \rightarrow \alpha + d\xi$ because,
as it is easy to see using,
for example, the Fourier expansion, such function $\xi$ cannot be
defined on the torus.

The existence of such zero modes is related to the
invariance of the action
${\cal S}(a)$ with respect to the transformation
$a \rightarrow a' = U_{c}a :=
a + c_{i} \alpha_{0i}$, where $\alpha_{0i}$ are zero modes. We
assume that they are normalized by $(\alpha_{0i}, \alpha_{0j}) =
\delta_{ij}$. According to the Hodge decomposition theorem (see,
for example, \cite{War})
\[
   {\cal H}^{p}(M) = Harm^{p}(M) \oplus (Harm^{p})_{\perp},
\]
where ${\cal H}^{p}(M)$ denotes the space of quadratically
integrable $p$-forms and $(Harm^{p})_{\perp}$ is the
orthogonal compliment of $Harm^{p}(M)$ with respect to the
scalar product (\ref{inner}). Let us insert the unit
\[
   1 = \int \prod_{i} dc_{i} \prod_{i} \delta \left( (\alpha_{0i},
U_{c} a) \right).
\]
After standard and simple manipulations we arrive to the expression
\beq
Z_{0}(\gamma) = \int \prod_{i} dc_{i}
\int_{{\cal A}^{(0)}} {\cal D}a \exp \left[ - \frac{1}{2}
(a, \Delta a) + i\nu e (J_{\gamma},a) \right]
\prod_{i} \delta \left((\alpha_{0i},a) \right) .
\label{Z03}
\eeq
Here we used the property of orthogonality
\[
(\alpha_{0i}, J_{\gamma}) = (\alpha_{0i}, \delta J_{\sigma}) =
(d \alpha_{0i}, J_{\sigma}) = 0,
\]
following from (\ref{Jg-JS}), and the fact that harmonic forms are
closed and co-closed.

On the space $(Harm^{p})_{\perp}$ the Laplacian $\Delta$
is invertible. Thus one can define the Green operator
$\tilde{\Delta}^{-1} \; : \; \Lambda^{p}(M) \rightarrow
(Harm^{p})_{\perp}$ as
\[
 \tilde{\Delta}^{-1} = \left( \Delta |_{(Harm^{p})_{\perp}}
 \right)^{-1} \circ \Pi_{p},
\]
where $\Pi_{p}$ is the projector of ${\cal H}^{p}(M)$ onto
$(Harm^{p})_{\perp}$ \cite{War}. In particular,
$\Pi_{1} = 1 - \sum_{i} \alpha_{0i} \alpha_{0i}^{*}$, where
\beq
  \left(\alpha_{0i} \alpha_{0i}^{*} \right) (\beta) = \alpha_{0i}
  \int \alpha_{0i} \wedge * \beta.   \label{a-a}
\eeq
The Green operator satisfies
\[
  \Delta \tilde{\Delta}^{-1} = \tilde{\Delta}^{-1} \Delta =
  \Pi_{1} \equiv 1 - \sum_{i}\alpha_{0i} \alpha_{0i}^{*}.
\]
Using this operator one can perform formally
the Gaussian functional integration in
(\ref{Z03}) and obtain the naive expression
\bea
Z_{0}(\gamma) & = & {\cal N} \exp \left[
- \frac{\nu^{2}}{2} e^{2} \Gamma (\gamma) \right],
\label{Z04}  \nonumber \\
\Gamma (\gamma) & = &  (J_{\gamma}, \tilde{\Delta}^{-1}
J_{\gamma}),   \label{Gamma}
\eea
where constant ${\cal N}$ accumulates all factors independent
of $\gamma$, they are irrelevant for our result. From
(\ref{Jg-delta}) it may seem that the inner product of the forms
in (\ref{Gamma}) is singular. We will show that in fact a well defined
value can be assigned to it. For this let us consider the following
regularized version of Eq. (\ref{Gamma}):
\beq
\Gamma (\gamma)  =   \lim _{\epsilon \rightarrow 0}
(J_{\gamma}, \tilde{\Delta}^{-1} J_{\gamma(\epsilon)}), \label{G-reg}
\eeq
where the closed path $\gamma(\epsilon)$ is some small deformation
of the closed path $\gamma$ which does not have common points with
$\gamma$ and $\gamma (0) = \gamma$ (we assume that the manifold
$M$ and the path $\gamma$ are such that such deformation is
possible). The product of two weak forms is not well
defined in general. However, we will see that in our case Eq.
(\ref{G-reg}) makes sense due to the facts that $J_{\gamma} =
\delta J_{\sigma}$, Eq. (\ref{Jg-JS}), and
$J_{\sigma} \in {\cal H}^{2}(M)$. We understand (\ref{G-reg}) as
\beq
\Gamma (\gamma)  =   \lim _{\epsilon \rightarrow 0}
(J_{\sigma}, d \tilde{\Delta}^{-1} \delta  J_{\sigma (\epsilon)}),
\label{G-reg1}
\eeq
with $\gamma (\epsilon) = \partial \sigma (\epsilon)$.

Now let us study the operator $d \tilde{\Delta}^{-1}
\delta$ and prove that it is equal to the projector
$\Pi_{2} \equiv 1 - w_{0i} w_{0i}^{*}$ on $L^{2}(M)$,
where $w_{0i}$ are harmonic 2-forms and $w_{0i}^{*}$ is the
functional corresponding to $w_{0i}$ in the sense explained
above (see Eq. (\ref{a-a})).
Indeed, since the Green operator commutes with all linear
operators which commute with the laplacian $\Delta$, in particular
$\delta  \tilde{\Delta}^{-1} = \tilde{\Delta}^{-1} \delta$ and
$ d \tilde{\Delta}^{-1} = \tilde{\Delta}^{-1} d $ \cite{War}, the
following chain of operator relations holds on $\Lambda^{2}(M)$:
\bea
  d \tilde{\Delta}^{-1} \delta & = &  d \delta \tilde{\Delta}^{-1} =
 (\Delta - \delta d)\tilde{\Delta}^{-1}   \nonumber \\
  & = & \Pi_{2} -  \delta \tilde{\Delta}^{-1} d = \Pi_{2}. \label{Dd3}
\eea
The last equality follows from the fact that the operator $d$ acts
trivially on $\Lambda^{2}(M)$ (recall that $\dim M = 2$).
The property (\ref{Dd3}) shows that the operator
$d \tilde{\Delta}^{-1} \delta$
is smooth enough and its action can be extended to
${\cal H}^{2}(M)$.

Let us return to the calculation of $\Gamma (\gamma)$ in Eq.
(\ref{G-reg1}). Using Eq. (\ref{Dd3}) we obtain that
\bea
\Gamma (\gamma) & = & \lim_{\epsilon \rightarrow 0} (J_{\sigma},
\Pi_{2} J_{\sigma(\epsilon)})  =
\left(J_{\sigma} ,(1- w_{0i} w_{0i}^{*})J_{\sigma}
\right) \nonumber \\
 & = & (J_{\sigma}, J_{\sigma}) - \left|(J_{\sigma}, w_{0i})
 \right|^{2}.            \label{GJ}
\eea

The form $\tilde{w}_{0} = \mu / \sqrt{V}$, where $\mu$ is
the volume 2-form (\ref{dv}), is the only (properly normalized)
zero mode (harmonic) form in $\Lambda^{2}(M)$.
Indeed, $d \tilde{w}_{0} = 0$ and using the properties of the
adjoint of the exterior derivative $\delta$ and of the Hodge $*$
operation we verify that
\[
\delta \tilde{w}_{0} = *d* \tilde{w}_{0} = * d \frac{1}{\sqrt{V}} = 0.
\]
This is in accordance with the fact that the space of harmonic
$p$-forms $Harm^{p}(M)$ is isomorphic to the cohomology group
$H^{p}(M;R)$ \cite{War}. In the case under consideration $p=2$,
$\dim M = 2$,
\[
Harm^{2}(M;R) \cong H^{2}(M;R) \cong R,
\]
$\dim Harm^{2}(M;R) = 1$, and, thus, all harmonic 2-forms are proportional to
$\tilde{w}_{0}$.

Using definition (\ref{JS-def}) we write
\beq
 (J_{\sigma}, \tilde{w}_{0}) = \int_{\sigma} \tilde{w}_{0} =
 \frac{1}{\sqrt{V}} \int_{\sigma} \mu
 = \frac{S}{\sqrt{V}}.   \label{GJ1}
\eeq
With the help of Eqs. (\ref{JS1}), (\ref{JS2})
we calculate the first term in (\ref{GJ}):
\beq
(J_{\sigma}, J_{\sigma}) = \int_{M} h_{\sigma}(x) \mu
\wedge * h_{\sigma}(x) \mu =
 \int_{M} h_{\sigma}^{2}(x) \mu =  \int_{\sigma} \mu = S. \label{GJ2}
\eeq
Combining formulas (\ref{Z04}), (\ref{GJ}) and results
(\ref{GJ1}), (\ref{GJ2}) we obtain the final expression for
the contribution of fluctuations around the monopoles:
\beq
Z_{0}(\gamma) = {\cal N} e^{-\frac{e^{2}}{2}\nu^2 \frac{S(V-S)}{V}}.
\label{Z05}
\eeq
Note that this expression is invariant under the transformation
$S \rightarrow (V-S)$, the same as $Z_{mon}(\gamma)$.
Using (\ref{Zmon1}) and (\ref{Z05}) we obtain the result
\[
Z(\gamma) = Z_{mon}(\gamma) Z_{0}(\gamma) =
  {\cal N}' \sum_{l=-\infty}^{\infty} \exp \left[ - \frac{e^{2}}{2} V
  \left( \frac{S}{V}\nu + l \right)^{2} - \frac{e^{2}}{2}\nu^2
  \frac{S(V-S)}{V} \right].
\]

The final expression for the expection value of the Wilson loop
variable for a homologically trivial loop $\gamma$ for the gauge group
$G=U(1)$ is
\beq
<T_{\gamma}> = \frac{
  \sum_{l=-\infty}^{\infty} \exp \left[ - \frac{e^{2}}{2} V
  \left( \frac{S}{V}\nu + l \right)^{2} -
  \frac{e^{2}}{2}\nu^2 \frac{S(V-S)}{V} \right]}{\sum_{l=
  -\infty}^{\infty} \exp \left[ - \frac{e^{2}}{2} V l^{2} \right]}.
  \label{Tg}
\eeq
It can be shown that for homologically non-trivial loops
$<T_{\gamma}> = 0$ (see also \cite{ALMMT}). Some discussion
of the properties of the expression (\ref{Tg}) will be given
at the end of Sect. 4 where it appears as a particular
case of the general formula for an arbitrary gauge group.

Let us analyze the quantity $E(e^{2}V,S/V) :=
-\ln < T_{\gamma}> $ which in a theory with fermions
characterizes the potential of the interaction.
$E(e^{2}V,S/V)$ can be read off from Eq. (\ref{Tg}) and
is plotted in Fig. 1 as a function of $S/V$ for two values of
$e^{2}V$ (for $\nu=1$).

\begin{figure}[ht]
\epsfxsize=0.9\hsize
\epsfbox{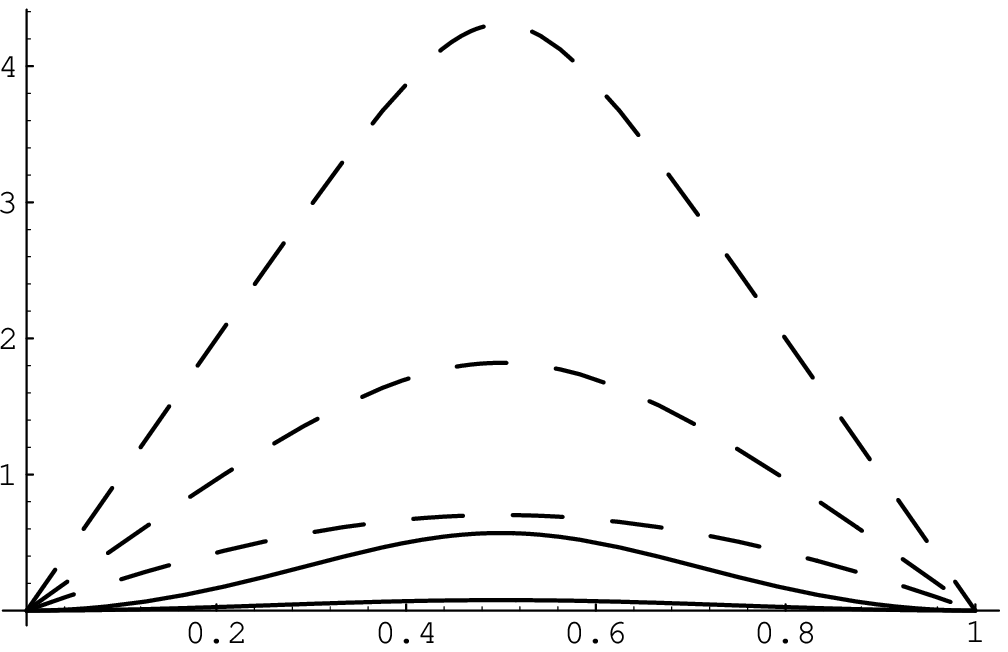}
\caption{ $E(e^{2}V,S/V)$ (dashed line) and $E_{mon}(e^{2}V,S/V)$
(solid line) as functions of $S/V$ for $e^{2} V=10$ (lower lines)
and $e^{2} V=20$ (upper lines) in the abelian gauge theory}
\end{figure}

In the same plot we also show the con\-tri\-bu\-ti\-on
$E_{mon}(e^{2}V,S/V)$ of the {\it abe\-li\-an} mo\-no\-poles
calculated from Eq. (\ref{Zmon}). We see from Fig. 1 (also from
the exact formula (\ref{Zmon1})) that the dependence of
$E_{mon}(e^{2}V,S/V)$ on $S/V$ at small $S/V$ is quadratic almost
till $S/V = 0.5$, where the curve reaches its maximum.
This, being combined with the contribution of the fluctuations,
gives the linear dependence (the area law) almost for
all $S/V$ in the interval $0 < S/V < 0.5$. The area law (linear
dependence on $S/V$) for the Wilson loop in the pure
Yang-Mills theory is considered as an indication of
the regime of confinement in the corresponding gauge
theory with quarks. Thus the quadratic behaviour of the monopole
contributions $E_{mon}(e^{2}V,S/V)$ seems to be an important feature
which gives rise to the linear dependence of the complete function.

In Sect. 5 we will study the analogous function for the contribution
of invariant connections.

\section{Lattice calculation of the Wilson loop for arbitrary gauge
group}

In this section we study a general pure gauge theory (\ref{action})
on a two-dimensional orientable compact manifold $M$.
We will calculate the functional integral (\ref{Zg-def}) again
but now for the case when the gauge group is an arbitrary compact
Lie group. As before we consider the case of a single simple loop
$\gamma$. We assume that $\gamma$ divides $M$ into two regions
$\sigma_1, \sigma_2$ of genera $r_1, r_2$ and areas $S_1, S_2$
respectively. The genus of $M$ is $r=r_1+r_2$ and its area
$V=S_1+S_2$ (see an example in Fig. 2).

\vskip 0.1cm
\begin{figure}[ht]
\epsfxsize=0.9\hsize
\epsfbox{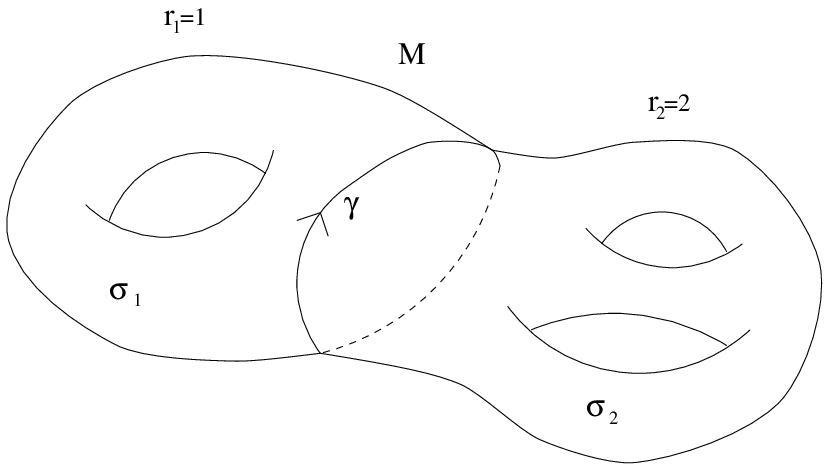}
\caption{A loop $\gamma$ on a surface $M$ with genus $r=3$.}
\end{figure}

Let us consider lattices on the manifold $M$ \cite{dz}. A lattice
can be viewed as a CW-complex $\Lambda =\Lambda_0
\cup \Lambda_1 \cup \Lambda_2$,
where $\Lambda_0,\Lambda_1$ and $\Lambda_2$ are finite sets
whose elements are 0-cells $s$ (sites), 1-cells $l$ (links) and
2-cells $p$ (plaquettes) respectively. For the purpose of computation
of  the expectation value of $T_\gamma$, we consider only lattices
compatible with $\gamma$, i.e. lattices such that $\gamma$ is
an 1-chain on them.
The incidence functions $I(s,l)$, $I(l,p)$ for pairs of
cells of correlative dimension take values 0, 1 or $-1$
if the first cell is not a face, is a positive face or is a
negative face of the second cell respectively. The topology
of a continuum surface, modelled by the lattice, is encoded
in the discrete structure given by $\Lambda$ and $I$.
The area of the plaquette $p$ is denoted by $|p|$.
The ``lattice spacing" $a$ is defined as the minimum length such
that every plaquette is confined in a circle of the radius $a$.

To calculate the functional integral (\ref{Zg-def}) we will
need a system of independent loops of the lattice $\Lambda$.
For this we fix a site $s_0$ of $\Lambda_0$, and for each plaquette
$p$ assign a connecting path $c_p$ from $s_0$ to $p$ and define
the loop $\gamma_p=c_p \partial p c_p^{-1}$. In addition we
choose a set of homotopically non-trivial loops
$a_i, b_i$ $i=1,\ldots ,r$ which play the role of the generators
of the first homotopy group of $M$ \cite{Bredon}, \cite{DNF},
\cite{ta}. The index $i$ labels the handles of $M$
(see Fig. 3 for an illustration).

\vskip 0.1cm
\begin{figure}[ht]
\epsfxsize=0.9\hsize
\epsfbox{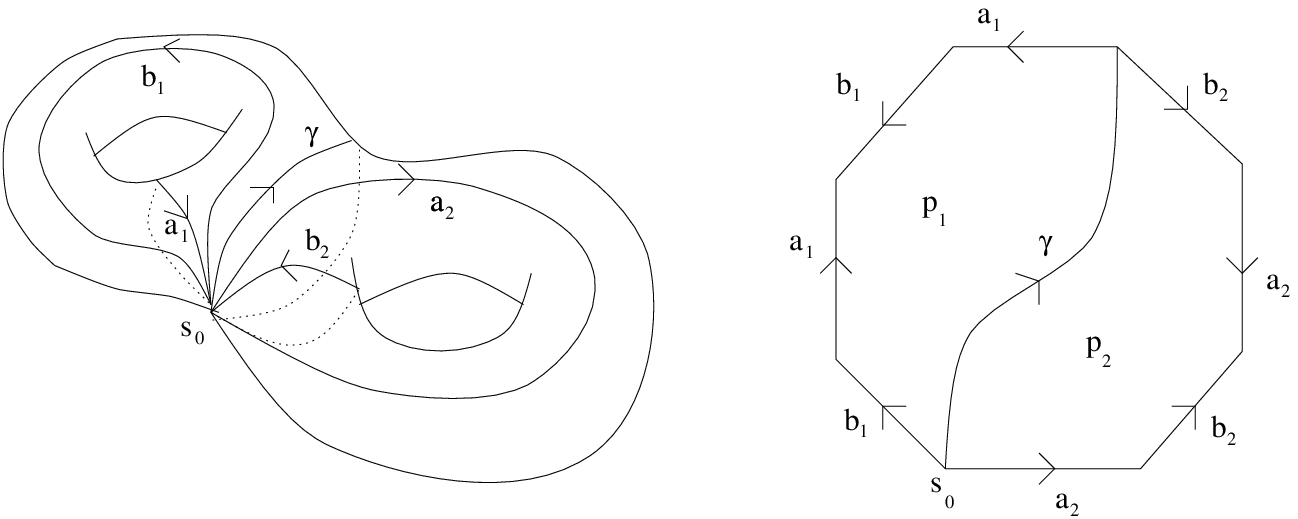}
\caption{
The system $\{\gamma_p,a_i,b_i\}$ on the 2-torus. The lattice
here is a minimal one with two plaquettes only. By cutting
the surface through the lines $a_i, b_i$ we obtain the
polygon of the figure. The relations are:
$p_2p_1=a_2b_2a_2^{-1}b_2^{-1}a_1b_1a_1^{-1}b_1^{-1}$
and $\gamma =p_1b_1a_1b_1^{-1}a_1^{-1}$.
}
\end{figure}

At this point it is useful to think of $M$ as being represented
by a polygon with certain identifications of its sides.
The system of loops $\{\gamma_p,a_i,b_i\}$ is not independent.
More precisely, it can be shown that there exists an appropriate
choice of paths $c_p$ and an ordering of the plaquettes so that
the following relation is true:
\beq
     \prod_p \gamma_p=\prod_{i} a_ib_ia_i^{-1}b_i^{-1}.
\label{homot}
\eeq
The idea of the construction is rather simple and is illustrated in
Fig. 4. The choice of the connecting paths $c_{p}$ is essentially
determined by the ordering of the plaquettes and is not unique. It
can be shown that final results do not depend on this ambiguity.

\vskip 0.1cm
\begin{figure}[ht]
\epsfxsize=0.9\hsize
\epsfbox{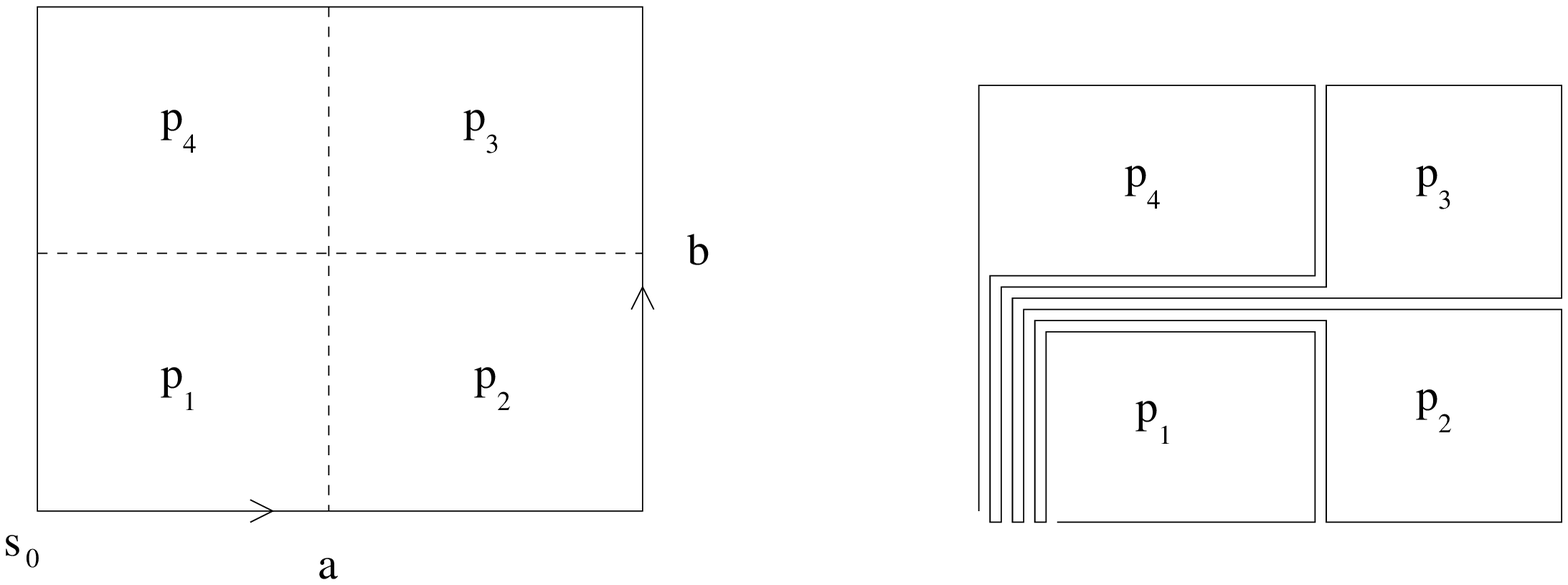}
\caption{A lattice with 4 plaquettes on a torus and
a choice of connecting paths such that
$aba^{-1}b^{-1}=\gamma_{p_1}\gamma_{p_2}\gamma_{p_3}\gamma_{p_4}$.
}
\end{figure}

Using the results of \cite{AL} (see also \cite{ALMMT}) one can give a
well-defined meaning to the heuristic measure ${\cal D}A $ in Eq.
(\ref{Zg-def}). It can be understood as the measure $d\mu_0$
constructed out of copies of the
Haar measure on the group $G$. This is possible if the action
can be written as a cylindrical function on the space
${\cal A} / {\cal T}$. The standard Yang-Mills action
${\cal S}(A)$ is not well defined on the lattice. Hence
one has to use some regularized action ${\cal S}_{\Lambda}(A)$
on the lattice $\Lambda$, calculate $Z_{\Lambda}(\gamma)$, which
is a lattice analog of (\ref{Zg-def}), and then take the limit
$Z(\gamma) = \lim_{a \rightarrow 0} Z_{\Lambda}(\gamma)$,
where $a$ is the lattice spacing of $\Lambda$.

In this article we consider a class of lattice actions with
the property
\beq
   e^{-{\cal S}_{\Lambda}(A)} = \prod _{p \in \Lambda}
  e^{-{\cal S}_{p}(H_{\gamma_p}(A))},
  \label{expact}
\eeq
where the product is taken over all plaquettes of $\Lambda$.

The function ${\cal S}_{p}$ is a real function over $G$
which satisfies the following three conditions:

1) ${\cal S}_{p}(g) = {\cal S}_{p}(g^{-1})$ for $g \in G$;

2) it reaches the absolute minimum on the identity element;

3) $\lim_{a \rightarrow 0}{\cal S}_{p}(H_{\gamma_p}(A))|p|^{-1} =
\frac{1}{2}Tr( F_{\mu \nu}(x))^{2}$.

Two main examples, which we have in mind, are:

a) the Wilson action \cite{Wi}:
\[
{\cal S}_{p}(g) = 1 - \frac{1}{d_F} Re \chi_F (g),
\]
where $F$ denotes the fundamental representation of $G$;

b) the heat-kernel action \cite{mo}:
\beq
{\cal S}_{p}(g) =-\ln \sum_R d_R \chi_R(g)
e^{-\frac{e^2}{2}c_2(R)|p|},         \label{hk}
\eeq
where $c_2(R)$ is the value of the Casimir operator for
the representation $R$. When $G=U(1)$ the heat-kernel action
is also called Villain action. With actions of this type
the integrand in (\ref{Zg-def}) is a cylindrical function.

The standard way to proceed would be the following.
Taking into account relation (\ref{homot}) we can
construct a system of independent loops $\beta_j$, $j=1,\ldots ,N$,
where $N=$(number of plaquettes)$+2r-1$. Let us introduce the notation
$g_{\beta_j} \equiv H_{\beta_j}(A)$, where the holonomy is defined
by Eq. (\ref{H-def}). The measure $d\mu_0$ is given by
$d\mu_0=\prod_{j=1}^{N} dg_{\beta_j}$, where $dg_{\beta_j}$ is the
Haar measure on $G$. Further, one expresses each loop
$\gamma_p$ or $a_i$, or $b_i$ in terms of $\beta_j$, like
$\gamma_p=\prod_{j}\beta_j$, so that
$H_{\gamma_p}(A)=\prod_{j}g_{\beta_j}$, and considers functions
\[
\nu_p(g_{\beta_1}, \ldots , g_{\beta_N}) =
\exp \left[ -{\cal S}_{p} \left( H_{\gamma_p}(A) \right) \right]
\]
on $G^{N}$ (direct product of $N$ copies of $G$). Since the lattice is
compatible with $\gamma$ we can also decompose $\gamma$ over the
independent loops $\beta_j$: $\gamma=\prod_{j}\beta_j$.

In this way one obtains that
\bea
Z(\gamma) & = & \lim_{a \rightarrow 0} Z_{\Lambda}(\gamma), \nonumber \\
 Z_{\Lambda}(\gamma) & \equiv & \int_{{\cal A}/{\cal T}} d \mu_{0}
   e^{- {\cal S}_{\Lambda}(A)} T_{\gamma}([A]) \label{ZL} \\
      & = & \int_{G^{N}}  \prod_j  dg_{\beta_{j}}
  \left( \prod _{p \in \Lambda} \nu_p(g_{\beta_1}, \ldots , g_{\beta_N})
  \right) \frac{1}{d_R} \chi_R \left(\prod_{\beta_{j} \in \gamma}
      g_{\beta_j} \right).   \nonumber
\eea
Here, as in Eq. (\ref{trh}), we assume that the Wilson loop
variable is in the irreducible representation $R$ of $G$.
Actually the integration in (\ref{ZL}) goes over a certain closure
$\overline{{\cal A} / {\cal T}}$, see details in Ref. \cite{ALMMT},
\cite{AL}.

In this article we will use another approach to the calculation
of integral (\ref{ZL}). We begin the calculation by changing it to
another form. The
variables $\{g_q\}\equiv\{g_{\gamma_p},g_{a_i},g_{b_i}\}$
will be treated as independent ones and relation (\ref{homot})
will be imposed by inserting the $\delta$-function
$$
\Delta [\{g_q\}]=
\delta_G (\prod_{p\in \Lambda} g_{\gamma_p} ,
\prod_i g_{a_i}g_{b_i}g_{a_i}^{-1}g_{b_i}^{-1} )
$$
where $\delta_G(g,h)$ is the Dirac distribution on the group $G$
with the main property
\[
\int_{G} dg \delta_G(g,h)f(g)=f(h).
\]
Then
\beq
 Z_{\Lambda}(\gamma) =
 \int_{G^{N+1}} (\prod _{i} dg_{a_i}dg_{b_i})
 \left( \prod _{p \in \Lambda} dg_{\gamma_p} e^{-{\cal S}_{p}
 (H_{\gamma_p}(A))} \right)\Delta [\{g_q\}]
 \frac{1}{d_R} \chi_R \left(
 \prod_{q \in \gamma} g_{q} \right).  \label{Zp-def}
\eeq
Now we use the fact that the action, given by Eq. (\ref{expact}),
can be expressed in terms of generalized Fourier coefficients.
We can write
\beq
e^{-{\cal S}_{p}(g)}=\sum_{R_{p}} d_{R_{p}} \chi_{R_{p}} (g)
B_{R_{p}}(p), \label{L-action1}
\eeq
where the coefficients $B_{R_{p}}(p)$ are equal to
\[
B_{R_{p}}(p)=\int dg e^{-{\cal S}_{p}(g)}\chi^{*}_{R_{p}}(g),
\]
and compute the functional integral (\ref{Zp-def}):
\bea
Z_{\Lambda}(\gamma ) & = &
  \int_{G^{N+1}} (\prod _{i} dg_{a_i}dg_{b_i})
  \left( \prod _{p \in \Lambda} dg_{\gamma_p}\right)  \nonumber \\
  & \times & (\prod _{p \in \Lambda}\sum_{R_p} d_{R_p}
 \chi_{R_p}(g_{\gamma_p})B_{R_p}(p)) \Delta [\{g_q\}]
 \frac{1}{d_R} \chi_R \left(\prod_{q \in \gamma} g_{q} \right)
               \nonumber \\
  & = & \int_{G^{N+1}} (\prod _{i} dg_{a_i}dg_{b_i})
  \left( \prod _{p \in \Lambda} dg_{\gamma_p}\right)  \nonumber \\
  & \times & (\sum_{\{R_p\}}\prod _{p \in \Lambda} d_{R_p}
  \chi_{R_p}(g_{\gamma_p})B_{R_p}(p)) \Delta [\{g_q\}] \frac{1}{d_R}
  \chi_R \left(\prod_{q \in \gamma} g_{q} \right)
             \nonumber \\
  & = & \sum_{\{R_p\}}(\prod_{p \in \Lambda} B_{R_p}(p))
\int_{G^{N+1}} (\prod _{i} dg_{a_i}dg_{b_i})
 \left( \prod _{p \in \Lambda} dg_{\gamma_p}\right) \nonumber \\
  & \times &  (\prod _{p \in \Lambda} d_{R_p} \chi_{R_p}(g_{\gamma_p}))
 \Delta [\{g_q\}] \frac{1}{d_R} \chi_R \left(\prod_{q \in \gamma}
 g_{q} \right).          \label{sumcol}
\eea
Here we interchanged the product over plaquettes and the
summation over representations, so that the last expression is a
sum over ``colorations'' of the surface, i.e. configurations or
sets $\{ R_{p} \}$ of irreducible representations of the gauge
group such that to each plaquette $q$ there is an
$R_{q} \in \{ R_{p} \}$ associated to it.
The integrals over $g_{\gamma_p}$ are non-zero only for
certain configurations, this depends on the
topology of the surface.

The decomposition of $\gamma$ is given in terms of
loops located in one of the regions whose border is $\gamma$,
say $\sigma_1$. Then,
\[
\gamma =\prod_p^{(1)}\gamma_p
[\prod_i^{(1)}a_ib_ia_i^{-1}b_i^{-1}]^{-1}
\]
where the superscript $(1)$ means the restriction to $\sigma_1$
(see Fig. 3).

To perform the $g_{\gamma_p}$-integration we use a procedure of
``lattice reduction". If plaquettes $p_1$ and $p_2$ share
a link which does not belong to $\gamma$ (in this case they are
in the same $\sigma_i$ component), then we shift the group
variable $g_{p_1}\rightarrow g_{p_1}g_{p_2}^{-1}$
and use the invariance of the measure. In this way $g_{p_2}$
disappears from the arguments of the $\delta$-function $\Delta$
and of $\chi_{R}$, and we can integrate it out obtaining
\beq
\int dg_{p_2} d_{R_{p_1}}\chi_{R_{p_1}} (g_{p_1}g_{p_2}^{-1})
 d_{R_{p_2}}\chi_{R_{p_2}} (g_{p_2})=d_{R_{p_1}}\chi_{R_{p_1}}
(g_{p_1})\delta_{R_{p_1} R_{p_2}}.
\label{gi1}
\eeq
The effect is that both plaquettes are forced to carry the same
representation in which case the common link is ``erased" and the two
plaquettes merge into one plaquette while the structure of
the integrand in (\ref{sumcol}) remains the same.
We continue integrating over the plaquette variables in this way
until we arrive to a lattice consisting only of two plaquettes
$P_1$ and $P_2$ in representations $R_1$ and $R_2$ correspondingly.
These two plaquettes can be envisioned by cutting the surface $M$
through the lines $\gamma$ and $a_i, b_i$.

Thus we obtain that
\beq
Z_\Lambda(\gamma )  =
 \sum_{R_1, R_2}\left( \prod_{p \in \sigma_1} B_{R_1}(p) \right)
 \left( \prod_{p \in \sigma_2} B_{R_2}(p) \right)
 \Gamma (\sigma_1, \sigma_2),       \label{zgam}
\eeq
where
\bea
\Gamma (\sigma_1, \sigma_2)  & = &
\int_{G^{2r+2}} (\prod _{i} dg_{a_i}dg_{b_i})
\left( dg_1dg_2\right)  d_{R_1}\chi_{R_1}(g_1) d_{R_2}\chi_{R_2}(g_2)
     \nonumber \\
  & \times & \delta_G (g_2g_1, \prod_{i} g_{a_i}g_{b_i}
  g_{a_i}^{-1}g_{b_i}^{-1}) \frac{1}{d_R} \chi_R
  \left( g_1[\prod_i^{(1)}g_{a_i}g_{b_i}g_{a_i}^{-1}
  g_{b_i}^{-1}]^{-1} \right).    \nonumber
\eea
The $g_2$-integration and the shift
$g_1\rightarrow (g_1\prod_i^{(1)}g_{a_i}g_{b_i}g_{a_i}^{-1}g_{b_i}^{-1})$
lead to
\bea
\Gamma (\sigma_1, \sigma_2) & = &
\int_{G^{2r+1}} dg_1(\prod _{i} dg_{a_i}dg_{b_i})
d_{R_1}\chi_{R_1} \left(g_1 \prod_i^{(1)}g_{a_i}g_{b_i}
g_{a_i}^{-1}g_{b_i}^{-1} \right)    \nonumber  \\
  & \times & d_{R_2}\chi_{R_2} \left( g_1^{-1}
 \prod_i^{(2)}g_{a_i}g_{b_i}g_{a_i}^{-1}g_{b_i}^{-1} \right)
        \frac{1}{d_R} \chi_R(g_1).  \nonumber
\eea
Using the property
\[
\int dg \chi_{R} (ghg^{-1}k)=d_R^{-1}\chi_{R}(h)\chi_{R}(k)
\]
and Eq. (\ref{gi1}) the integration over $a_i, b_i$ can be fulfilled,
and one obtains
\bea
& & \int_{G^{2r_1}} (\prod _{i}^{(1)} dg_{a_i}dg_{b_i})
 d_{R_1}\chi_{R_1} \left( g_1 \prod_i^{(1)}g_{a_i}g_{b_i}
 g_{a_i}^{-1}g_{b_i}^{-1} \right)   \nonumber \\
 & = & \int_{G^{r_1}} (\prod _{i}^{(1)} dg_{a_i})
 d_{R_1}d_{R_1}^{-r_1}\chi_{R_1} \left(
 g_1 \prod_i^{(1)}g_{a_i} \right)
 \prod_i^{(1)}\chi_{R_1} \left( g_{a_i}^{-1} \right)   \nonumber \\
 & = & d_{R_1}^{1-2r_1}\chi_{R_1}(g_1),  \nonumber
\eea
so that
\bea
\Gamma (\sigma_1, \sigma_2)  & = &
\int_{G} dg_1 d_{R_1}^{1-2r_1}\chi_{R_1}(g_1)
 d_{R_2}^{1-2r_2}\chi_{R_2}(g_1^{-1})
 \frac{1}{d_R} \chi_R (g_1)   \nonumber \\
 & = & \frac{1}{d_{R}} d_{R_1}^{1-2r_1}d_{R_2}^{1-2r_2}\{ R_1,R,R_2 \},
\label{gam}
\eea
where $\{R_1,R,R_2\}$ is the number of times the representation
$R_2$ is contained in $R_1\otimes R$ (fusion number).

Substituting (\ref{gam}) into (\ref{zgam}) we obtain the final
expression
\[
Z_\Lambda(\gamma )  =
 \sum_{R_1, R_2}(\prod_{p \in \sigma_1} B_{R_1}(p))
                 (\prod_{p \in \sigma_2} B_{R_2}(p))
        \frac{1}{d_{R}} d_{R_1}^{1-2r_1}d_{R_2}^{1-2r_2}\{R_1,R,R_2\}.
\]
The vacuum expectation value of the Wilson loop variable in the
two-dimensional gauge theory on the lattice is equal to
\[
<T_\gamma >_{\Lambda}  =
\frac{ \sum_{R_1, R_2} \left( \prod_{p \in \sigma_1} B_{R_1}(p) \right)
 \left( \prod_{p \in \sigma_2} B_{R_2}(p) \right)
        d_{R_1}^{1-2r_1}d_{R_2}^{1-2r_2} \{R_1,R,R_2\} }
{d_{R}\sum_{R_3} \left( \prod_{p \in \Lambda} B_{R_3}(p) \right)
d_{R_3}^{2-2r}}.
\]

To obtain the continuum limit one should, in general,
compute the coefficients $B_{R}(p)$ and take their
infinitesimal area limit. This is the case if we work, for example,
with the Wilson action. If we use instead the heat-kernel action
(\ref{hk}) we have
\[
B_{R}(p) = e^{-\frac{e^2}{2}c_2(R)|p|},
\label{bhk}
\]
and the vacuum expectation value
\beq
<T_\gamma >_{\Lambda}  =
\frac{\sum_{R_1, R_2} \tilde{B}_{R_1}^{S_1}
                 \tilde{B}_{R_2}^{S_2}
        d_{R_1}^{1-2r_1}d_{R_2}^{1-2r_2}\{R_1,R,R_2\}}
{d_{R}\sum_{R_3}\tilde{B}_{R_3}^{V}d_{R_3}^{2-2r}},
\label{Tfin}
\eeq
where $\tilde{B}_{R} = \exp (-e^2c_2(R)/2)$, is independent of the
lattice spacing, so that it gives already the continuum limit.
Of course, as one can check, in the limit $a\rightarrow 0$ the
results of the calculations with any action of the type
(\ref{expact}) coincide (see \cite{ALMMT} for examples).

\vspace{5mm}

Let us now consider the above result in the abelian case.
For abelian groups $d_{R}=1$ and the topology of
the surface is irrelevant. For $G=U(1)$ the irreducible
representations are labelled by integers $n\in Z$.
If the region $\sigma_1$ carries $R_1=n$, then
from the definition of $\{R_1,R,R_2\}$ in Eq. (\ref{gam})
$R_2=n+\nu$, where $\nu$ labells the irreducible representation
for which the Wilson loop variable is calculated (see (\ref{Tnu})).
Since $B_n = \exp \left(-e^2n^2/2 \right)$, we have that
\[
<T_{\gamma} >= \frac{\sum_{n=-\infty}^{\infty} B_{n}^{S_1}
B_{n+\nu}^{S_2}}
{\sum_{n=-\infty}^\infty B_n^V}
=\frac{\sum_{n=-\infty}^\infty e^{-S_1e^2n^2/2}e^{-S_2e^2(n+\nu)^2/2}}
{\sum_{n=-\infty}^\infty e^{-Ve^2n^2/2}}.
\]
Denoting $S_1$ and $S_2$ as $S$ and $V-S$ we obtain
the following expression
\beq
<T_{\gamma}> = \exp \left\{ - \frac{e^2 V \nu^2}{2} \frac{S}{V}
 \left( 1 - \frac{S}{V} \right)  \right\}
\frac{
\sum_{n=-\infty}^{\infty} e^{-e^2 V (n+S \nu /V)^2/2}}
{\sum_{n=-\infty}^{\infty} e^{-e^2 V n^2/2}},
\label{Tlat}
\eeq
which coincides with the result (\ref{Tg}) in Sect. 3

Another derivation of the formula for $<T_{\gamma }>$ which uses
the differential calculus of the lattice is given in the
Appendix.

\vspace{5mm}

A few comments are relevant here:

\begin{itemize}
\item Eq. (\ref{Tlat}) is obviously invariant
under the transformation $S \leftrightarrow V-S$.

\item Although we have considered compact surfaces only,
the same technique applies to the non-compact case.
The main difference is that in the non-compact case one of the regions,
say, $\sigma_{1}$ is infinite and the irreducible representation
corresponding to it has to be trivial. This case can also
be obtained as the limit $V\rightarrow \infty$
of the compact one. Performing the limit in Eq. (\ref{Tlat}) we see
that the expectation value of the Wilson loop functional for
the Euclidean plane $R^2$ or any non-compact
surface has the typical area law behaviour,
\beq
<T_{\gamma}>=e^{-\frac{e^2}{2}\nu^2S},   \label{T-Vinf}
\eeq
where $S$ denotes the area of the region surrounded by the
loop $\gamma$.

\item For finite $V$ the same behaviour is obtained in
the strong coupling limit $e^2 V \gg 1 $.
That is, in this limit the region with the smaller area dominates.

\item In the weak coupling limit $e^2 V \ll 1 $
\[
<T_{\gamma }> \sim  \exp \left\{ - \frac{(e^2 V) \nu^2}{2} \frac{S}{V}
 \left( 1 - \frac{S}{V} \right)  \right\}.
\]
This behaviour can be deduced from the expression
\[
<T_{\gamma }> = \exp \left\{ - \frac{(e^2 V) \nu^2}{2} \frac{S}{V}
 \left( 1 - \frac{S}{V} \right)  \right\}
\frac{ 1 + 2 \sum_{l=1}^\infty e^{-\frac{2\pi^2}{e^2 V}l^2}
\cos (2\pi l \frac{S}{V}\nu ) }
{\sum_{l=-\infty}^\infty e^{-\frac{2\pi^2}{e^2 V }l^2}}
\]
which follows from Eq. (\ref{Tlat}) after applying the Poisson
summation formula (\ref{Poisson}). This coincides with
the contribution (\ref{Z05}) of the trivial sector with zero
topological charge, as it can be expected, since in the weak coupling
limit the contributions of the monopoles are exponentially supressed,
see Eq. (\ref{Zmon}).
\end{itemize}

\vspace{5mm}

For non-abelian groups properties of the representations
of the particular group play an explicit role in (\ref{Tfin}).
As an example let us consider the case $G=SU(2)$ with $\gamma$ in
the fundamental representation. We label the irreducible
representations by $j=0,\frac{1}{2},1,\ldots $
as usual. Then $d_j=2j+1$ and $c_2(j)=j(j+1)$, and the vacuum
expectation value of the Wilson loop variable is equal to
\[
<T_{\gamma }> = \exp \left\{ -\frac{e^2 V}{8} \frac{S}{V}
 \left( 1 - \frac{S}{V} \right) \right\}
\frac{ \sum_{m=-\infty; m \neq 0,-1}^{\infty} m^{1-2r_1}(m+1)^{1-2r_2}
e^{-\frac{e^2 V}{8} \left( m+\frac{S}{V} \right)^2}}
{\sum_{m=-\infty; m \neq 0}^{\infty} m^{2-2r}e^{-\frac{e^2 V}{8} m^2}}.
\]

\section{Wilson loop variables for invariant connections.}

In Sect. 3, while calculating the vacuum expectation average of
the Wilson loop functional
in the abelian case, we studied the contribution of instantons.
In the present section we
will calculate Wilson loop variables for a special class of connections
called invariant connections.
The result of the calculation will be compared with the exact formula
derived above, and this will allow
us to understand how much of the information is captured by the
invariant connections. Our interest
in this class of connections is motivated by the fact that the
analogous calculation can be
carried out in Yang-Mills theory in any dimension provided,
of course, that non-trivial
invariant connections exist.

First we discuss the definition of invariant connections and
their construction.

Invariant connections in a principal fibre bundle are
connections which are invariant under the action of a group
of transformations on the bundle. In field theory models this
situation occurs when some group $K$ acts on
the space-time manifold (the base of the fibre bundle) and its action
can be lifted to the action of a subgroup of the group of automorphisms.
Then $K$ is called symmetry group. Let $P(M,G)$ be a principal
fibre bundle with the structure group $G$ over the base $M$
and $L_{k}$, $k \in K$ be an action of the symmetry group in $P$.
Since $K$ acts as a subgroup of the group of automorphisms of $P$,
the following relation holds:
\[
   (L_{k} p)g = L_{k} (pg) \; \; \; \mbox{for all} \; \;
   k \in K, \; g \in G \;
  p \in P,
\]
where multiplication by an element of $G$ from the right denotes the
canonical action of the structure group on the principal fibre bundle.
Let us denote by
$O_{k}$ the corresponding action of the symmetry group on the base
manifold. It is defined in a natural way by
\beq
      O_{k} \left( \pi (p) \right) =  \pi \left( L_{k} p \right),
      \; \; \; k \in K, \; \;
     p \in P,  \label{O-L}
\eeq
where $\pi$ is the canonical projection in $P$.
For our purposes it is enough to consider
the case when the action of $K$ on $M$ is transitive. More general
situation, when $M$ consists
of orbits of the action of $K$ occurs, for example, in the
problem of the coset space dimensional reduction
\cite{Manton} (see Ref. \cite{KMRV} for a review).

Now we are ready to give the definition. A connection in $P$ is said
to be invariant
with respect to transformations of the group $K$ if its connection
form $w$ satisfies
\beq
      L_{k}^{*} w = w     \label{inv-def}
\eeq
for all $k \in K$. In case of gauge theory the relevant question
is what is the property of the gauge potential given by an invariant
form. Such potentials were introduced in Refs. \cite{Schwarz} and
were called symmetric potentials. Initially
they were used for the
construction of Ans\"atze for solutions of equations of motion.
It can be shown that condition (\ref{inv-def}) implies that
for any $k \in K$ there exists a gauge tranformation $g_{k}(x) \in G$
such that the gauge potential $A_{\mu}$, corresponding to
the invariant connection form $w$, satisfies
\beq
    (O_{k} A)_{\mu} = g_{k}(x)^{-1} A_{\mu}(x) g_{k}(x) +
    \frac{1}{i e} g_{k}(x)^{-1} \partial_{\mu}
   g_{k}(x)   \label{symm-pot}
\eeq
for all $k \in K$, where the l.h.s. is the field obtained by
the space-time transformation. This formula means that the
symmetric potential is invariant under transformations from
$K$ up to a gauge transformation.

Invariant connections were first studied by Wang \cite{Wang} who
proved a theorem which
gives a complete characterization of such connections (see \cite{KN}).
We will follow Ref. \cite{KMRV} in the description of this construction.

The fact that the symmetry group acts transitively on the base
means that $M$ is a coset space $K/H$, where $H$ is a subgroup
of $K$ called the isotropy group. $K$ acts on $K/H$ in the canonical
way: $O_{k} [k_{1}] = [k k_{1}]$, where points $x \in M$ are
understood as classes $x = [k] \equiv kH$ of $K/H$.
Then the origin $o$ of $M = K/H$, that is the
class containing the unit of $K$, $o=[e]=H$, is stable under
the action of the elements from $H$. It follows immediately
from Eq. (\ref{O-L}) that transformations $L_{h}$ with $h \in H$ act
vertically on $P$. Thus, for each $p \in P$ there exists a mapping
$\chi_{p}: H \rightarrow G$ defined by $L_{h} p = p \: \chi_{p}(h)$
for any $h \in H$. It can be shown that this mapping is in fact
a group homomorphism. The existing symmetry allows to carry out the
reduction of the initial fibre bundle to its
subbundle over the origin $o$ such that the homomorphisms $\chi_{p}$
are the same for
points of this subbundle (we will denote them by $\chi$). Invariant
connection forms
$w$ in $P$ are characterized by their values in this subbundle and
are in a one-to-one correspondence with some mapping $\phi$ which
we will describe right now. Let ${\cal G}$,
${\cal K}$ and ${\cal H}$ be the Lie algebras of the groups $G$, $K$
and $H$ respectively.
If $H$ is a closed compact subgroup of $K$, the case we have in
mind, then the homogeneous space $K/H$ is reductive, i.e.
exists the decomposition
\beq
    {\cal K} = {\cal H} + {\cal M}     \label{K-decomp}
\eeq
with $ad ({\cal H}) {\cal M} = {\cal M}$,
where $ad$ denotes the adjoint action of the group on its Lie
algebra. The mapping $\phi$ maps from ${\cal M}$ into ${\cal G}$
and is equivariant, i.e. satisfies the condition
\beq
      \phi \left( ad (h) X \right) = ad (\chi (h)) \phi(X)
      \label{equiv}
\eeq
for all $h \in H$ and $X \in {\cal M}$. An explicit formula for an
invariant form in terms of $\chi$ and $\phi$ can be also given. For
our purposes it is more convenient to
write such formula for the pull-back of $w$ to the base $M$ with
respect to a (local) section
$s$. Let us denote such form on $M$ by $A$, $A = s^{*} w$.
Of course, if $w$ is the
invariant form, then $A = A_{\mu} dx_{\mu}$, where $A_{\mu}$ is the
corresponding symmetric
potential satisfying (\ref{symm-pot}). Let $\theta$ be the canonical
left-invariant 1-form on the Lie group $K$
with values in ${\cal K}$, and $\bar{\theta}$ is its pull-back
to $K/H$. If $k(x) \in K$
is a representative of the class $x \in K/H$, then
$\bar{\theta} = k(x)^{-1} dk(x)$.
We decompose the 1-form $\bar{\theta}$ into the ${\cal H}$-
and ${\cal M}$-components in accordance with (\ref{K-decomp}):
$\bar{\theta} = \bar{\theta}_{\cal H} + \bar{\theta}_{\cal M}$.
Then it can be shown that for a given homomorphism $\chi$ all
invariant connections are described by
\beq
        A = \frac{1}{ie} \left( \chi (\bar{\theta}_{\cal H}) +
        \phi (\bar{\theta}_{\cal M}) \right),   \label{A-inv}
\eeq
where the mapping $\phi$ satisfies the equivariant condition
(\ref{equiv}). Here we denoted by the same letter $\chi$ the
homomorphism from ${\cal H}$ to ${\cal G}$ induced by the group
homomorphism. Eq. (\ref{A-inv}) is essentially the result of
the Wang theorem \cite{KN}.

We see that the problem of construction of $K$-invariant
connections on a homogeneous space $M=K/H$ is reduced to
the construction of the mappings $\phi: {\cal M} \rightarrow
{\cal G}$. For this we need to solve the constraints
(\ref{equiv}). For our purpose it is more convenient to consider
the same constraint in the infinitesimal form
\beq
      \phi \left( Ad (Y) X \right) = Ad (\chi (Y)) \phi(X), \; \;
      Y \in {\cal H}, \; \;
      X \in {\cal M},    \label{equiv1}
\eeq
where $Ad$ denotes the adjoint action of the algebra. Due
to the algebraic structure
inherited by the vector subspace ${\cal M} \subset {\cal K}$
$Ad (h)$ is given by the Lie algebra bracket: $Ad(Y)X = [Y,X]$.
Similarly, if $X_{1}, X_{2} \in {\cal G}$,
$Ad(X_{1}) X_{2} = [X_{1}, X_{2}]$, where now the brackets denote
the multiplication in the Lie algebra ${\cal G}$. This case
takes place in the r.h.s. of Eq. (\ref{equiv1}).

An effective technique for solving constraint (\ref{equiv1})
was developed in Refs. \cite{VK}
(see also \cite{KMRV}). The main idea is to consider condition
(\ref{equiv1}) as the intertwining condition and the mapping $\phi$ as
the intertwining operator which intertwines representations of
${\cal H}$ in the the vector spaces ${\cal M}$ and ${\cal G}$.
Then the general structure of $\phi$ is given by Schur's lemma.
One has to decompose the vector spaces ${\cal M}$ and ${\cal G}$
into irreducible representations with respect to the actions
$Ad ({\cal H})$ and
$Ad ( \chi ({\cal H}))$ of the algebra ${\cal H}$ respectively
and choose pairs of equivalent
representations. In practice it is easier to work not with the
original compact Lie algebras and
subspaces but with their complexified versions. In this case Schur's
lemma tells that the
intertwining operator $\phi$ is the identity between the subspaces
carrying equivalent
irreducible representations of ${\cal H}$ and is zero between the
subspaces carrying
non-equivalent ones.

As concrete examples we will consider the Yang-Mills theory on
the two-dimensional sphere $S^{2}$ with the gauge groups $G=U(1)$,
$SU(2)$ and $SO(3)$.
The sphere is realized as a coset space $S^{2} = SU(2)/U(1)$.
First let us construct
the 1-forms $\bar{\theta}_{\cal H}$ and $\bar{\theta}_{\cal M}$
which appear in Eq. (\ref{A-inv}). As in Sect. 3 we introduce
the angles $\vartheta$ and $\varphi$, parametrizing the sphere,
and the neighbourhoods $U_{1}$ and $U_{2}$ described by
(\ref{chart1}) and (\ref{chart2}). As generators of $K=SU(2)$ we take
$Q_{j}=\tau_{j}/2$ ($j=1,2,3$), where $\tau_{j}$ are the Pauli
matrices. Let the subgroup $H=U(1)$ be generated by
$Q_{3}$. Then the one-dimensional algebra ${\cal H}$ is spanned by
$Q_{3}$, and the vector space ${\cal M}$ in (\ref{K-decomp})
is spanned by $Q_{1}$ and $Q_{2}$.
Consider now the decomposition of the algebra ${\cal K}$,
which is the complexification of the Lie algebra of the group
$K=SU(2)$. We denote by $e_{\alpha}$ and
$e_{-\alpha}$ the root vectors and by $h_{\alpha}$ the corresponding
Cartan element of this algebra and take
\[
   e_{\pm \alpha}= \tau_{\pm} = \frac{1}{2}(\tau_{1}\pm i\tau_{2}),
   \; \; \; h_{\alpha} = \tau_{3}
\]
with
\[
    Ad \left( h_{\alpha} \right) (e_{\pm \alpha}) =
    [h_{\alpha}, e_{\pm \alpha}] =
   \pm 2 e_{\pm \alpha}.   \label{h-e}
\]
Then ${\cal H}= C h_{\alpha}$,
${\cal M} = C e_{\alpha} + C e_{-\alpha}$ and
the decomposition of the vector space ${\cal M}$ into
irreducible invariant subspaces of ${\cal H}$
is described by the following decomposition of the
representations:
\beq
    \underline{2} \rightarrow (2) + (-2),  \label{M-decomp}
\eeq
where in the r.h.s. we indicated the eigenvalues of $Ad (h_{\alpha})$,
and the space ${\cal M}$ of reducible representation is
indicated by its dimension in the l.h.s.

We choose the local representatives $k^{(j)}$ $(j=1,2)$ of
points of the neighbourhood $U_{j}$ of the coset space $S^{2} =
SU(2)/U(1)$ as follows
\[
  k^{(1)}(\vartheta, \varphi) = e^{-i \varphi \frac{\tau_{3}}{2}}
  e^{i \vartheta \frac{\tau_{2}}{2}}
  e^{i\varphi \frac{\tau_{3}}{2}},
  \; \; \;
  k^{(2)}(\vartheta, \varphi) = e^{i\varphi \frac{\tau_{3}}{2}}
  e^{i(\vartheta-\pi) \frac{\tau_{2}}{2}}
  e^{-i\varphi \frac{\tau_{3}}{2}}.
\]
The functions $k^{(i)}: U_{i} \rightarrow SU(2)$ can also be viewed as
local sections of the principal fibre bundle
$K = P(K/H,H)$ over the base $K/H=S^{2}$ with the structure group
$H=U(1)$. By straightforward
computation one obtains the forms $\bar{\theta}_{\cal H}$ and
$\bar{\theta}_{\cal M}$:
\bea
    \bar{\theta}^{(i)} & = & \left(k^{(i)}\right)^{-1} d k^{(i)} =
    \bar{\theta}_{\cal H}^{(i)} + \bar{\theta}_{\cal M}^{(i)},
                                           \nonumber \\
    \bar{\theta}_{\cal H}^{(1)} & = & i \frac{\tau_{3}}{2}
    \left(1 - \cos \vartheta \right) d\varphi,   \label{theta1} \\
    \bar{\theta}_{\cal M}^{(1)} & = &
    i \frac{\tau_{1}}{2} \left(-\sin \varphi d\vartheta -
    \sin \vartheta \cos \varphi d\varphi \right) +
       i \frac{\tau_{2}}{2} \left( \cos \varphi d\vartheta -
       \sin \vartheta \sin \varphi  d\varphi \right),
                                  \nonumber \\
    \bar{\theta}_{\cal H}^{(2)} & = & -i \frac{\tau_{3}}{2}
    \left(1 + \cos \vartheta \right)
    d\varphi,    \label{theta2} \\
    \bar{\theta}_{\cal M}^{(2)} & = &
       i \frac{\tau_{1}}{2} \left(\sin \varphi d\vartheta -
       \sin \vartheta \cos \varphi d\varphi \right) +
       i \frac{\tau_{2}}{2} \left( \cos \varphi d\vartheta +
       \sin \vartheta \sin \varphi d\varphi \right).
                                  \nonumber
\eea
Before using the general formula (\ref{A-inv}) for the invariant gauge
connections one has to specify the gauge group and the embedding
$\chi(H) \subset G$. We consider three examples.

\noindent \underline{Example 1. $G=U(1)$.} Let us realize elements
of $G$ by unimodular complex numbers $\exp (it)$, where $t$ is real.
The group homomorphisms $\chi: H \rightarrow G$ are labelled by
integers $n$ and given by
\[
    \chi_{n} \left( e^{-i\frac{\tau_{3}}{2}t} \right) =
    e^{-i\frac{n}{2}t}.
\]
The corresponding algebra homomorphisms are determined by
$\chi_{n}(h_{\alpha}) = n$. The one-dimensional
vector space ${\cal G}=Lie(G)$ is the space of an
irreducible representation of $\chi ({\cal H})$ characterized
by zero eigenvalue of $Ad ( \chi (h_{\alpha}))$. Taking
decomposition (\ref{M-decomp}) of ${\cal M}$ into account it
is easy to see that intertwining condition (\ref{equiv1})
fulfills only for $\phi=0$. Thus, all invariant connections are
labelled by the integer $n$ and the gauge potential is given by
\bea
     A^{(i)}_{n} & = & \frac{1}{ie} \chi_{n}
     (\bar{\theta}_{\cal H}^{(i)}),
     \; \;
       i = 1,2; \nonumber \\
     A^{(1)}_{n} & = & \frac{n}{2e} (1 - \cos \vartheta) d\varphi,
     \nonumber \\
    A^{(2)}_{n} & = & -\frac{n}{2e} (1 + \cos \vartheta) d\varphi,
    \nonumber
\eea
where the lower indices of $A^{(1)}_{n}$ and $A^{(2)}_{n}$ correspond
to the labels of the homomorphism $\chi_{n}$.
These are the expressions for the Dirac-Wu-Yang abelian
monopole (\ref{U1-mon}), and $n$ characterizes the monopole charge.
Monopoles with different $n$ are described by connections
in non-equivalent fibre bundles. In accordance
with our discussion in Sect. 2 ${\cal B}_{U(1)}(S^{2}) \cong Z$, i.e.
such bundles are indeed classified by integers.
We have shown that all of them appear as particular invariant
connections.

\noindent \underline{Example 2. $G=SU(2)$.}
Let $E_{\alpha}$, $E_{-\alpha}$ and $H_{\alpha}$ be the root
vectors and the Cartan element of the algebra ${\cal G} = A_{1}$,
which appears as complexification of the Lie algebra of $G$.
We assume that they are given by the same combinations of the
Pauli matrices as the corresponding elements of complexified
Lie algebra of $K$ described above. The group homomorphism
$\chi: H=U(1) \rightarrow G=SU(2)$ is given by the expression
\beq
   \chi \left( e^{i\frac{\tau_{3}}{2} \alpha_{3}} \right) =
e^{i \kappa \frac{\tau_{3}}{2} \alpha_{3}} =
 \cos (\kappa \frac{\alpha_{3}}{2})
  + i \tau_{3} \sin (\kappa \frac{\alpha_{3}}{2}),  \label{tau-hom}
\eeq
and it is easy to check that this definition is consistent if
$\kappa$ is integer. Therefore the homomorphism is
again labelled by $n \in Z$. The induced algebra homomorphism
is given by
\beq
     \chi_{n} (h_{\alpha}) = n H_{\alpha}.  \label{SU2-hom}
\eeq
The three-dimensional space ${\cal G}$ of
the adjoint representation of $A_{1}$ decomposes into three
1-dimensional irreducuble invariant subspaces of
$\chi_{n}({\cal H})$ and the decomposition
is characterized by the following decomposition of representations:
\beq
    \underline{3} \rightarrow (0) + (2n) + (-2n)  \label{G-decomp}
\eeq
(in brackets we indicate the eigenvalues of
$Ad (\chi_{n} ( h_{\alpha}))$).

Let us now compare decompositions (\ref{M-decomp}) and
(\ref{G-decomp}). For $n \neq \pm 1,0$ there are no equivalent
representations in the decomposition of ${\cal M}$ and ${\cal G}$ and
the intertwining operator $\phi: {\cal M} \rightarrow {\cal G}$
is zero. It also turns out to be zero for $n=0$. In these cases
according to (\ref{A-inv})
\bea
A^{(1)}_{n} & = &  \frac{n}{2e} \tau_{3} (1 - \cos \vartheta)
  d \varphi,    \nonumber \\
A^{(2)}_{n} & = &  -\frac{n}{2e} \tau_{3} (1 + \cos \vartheta)
   d \varphi. \label{A-SU2}
\eea

If $n=1$ or $n=-1$ the results are more interesting.
Let us consider the case $n=1$ first. Comparing (\ref{M-decomp})
and (\ref{G-decomp}) we see that there are pairs of representations
with the same eigenvalues and, therefore, the intertwining
operator $\phi$ is non-trivial. It is determined by its action
on the basis elements of ${\cal M}$:
\beq
   \phi (e_{\alpha}) = f_{1} E_{\alpha}, \; \;
    \phi (e_{-\alpha}) = f_{2} E_{-\alpha},  \label{SU2-phi}
\eeq
where $f_{1}$, $f_{2}$ are complex numbers. The fact that the initial
groups and algebras are compact implies a reality condition \cite{KMRV}
which tells that $f_{1} = f_{2}^{*}$. Thus, the operator
$\phi$ and the invariant connection are parametrized by one
complex parameter $f_{1}$ (we will suppress its index from now on).
Using Eqs. (\ref{theta1}), (\ref{theta2}), (\ref{SU2-hom})
and (\ref{SU2-phi}) we obtain from (\ref{A-inv}) that
\beq
A^{(1)}_{1} = \frac{1}{2e} \left( \begin{array}{cc}
 (1 - \cos \vartheta) d\varphi & f e^{-i\varphi}
 (-id\vartheta - \sin \vartheta d \varphi) \\
f^{*}e^{i\varphi}(id\vartheta - \sin \vartheta d \varphi) &
 -(1 - \cos \vartheta) d\varphi
     \end{array}  \right),    \label{A-SU2-1}
\eeq
\beq
A^{(2)}_{1} = \frac{1}{2e} \left( \begin{array}{cc}
-(1+\cos \vartheta) d\varphi & f e^{i\varphi} (-id\vartheta -
\sin \vartheta d \varphi) \\
f^{*}e^{-i\varphi}(id\vartheta - \sin \vartheta d \varphi) &
(1+\cos \vartheta) d \varphi
     \end{array}  \right).  \label{A-SU2-2}
\eeq
The curvature form $F = dA + \frac{ie}{2} [A,A]$ is described by
the unique expression on the whole sphere and is equal to
\beq
    F_{n} = - \frac{1}{2e} \tau_{3} \left( |f|^{2} - 1 \right) \sin
    \vartheta  d\vartheta \wedge d\varphi.      \label{F-SU2}
\eeq
The action for such configuration is equal to
\beq
   {\cal S}_{inv} (f) = \frac{\pi}{2 e^{2}} \frac{1}{R^{2}}
    \left(|f|^{2} - 1 \right)^{2},     \label{S-SU2}
\eeq
where $R$ is the radius of the sphere. Due to the $K$-invariance
any extrema of the action found
withing the subspace of invariant connections is also an
extremum in the space of all connections \cite{KMRV}. From Eq.
(\ref{S-SU2}) we see that there are two types of extrema in
the theory: the maximum at $f=0$ and the minima at
$f$ satisfying $|f|=1$. Only one of them, the trivial
extremum, was found in Ref. \cite{RP} as a spontaneous
compactification solution in six-dimensional Kaluza-Klein theory.

Similar situation takes place for $\kappa=-1$. Again there exists
a 1-parameter family of invariant connections parametrized
by a complex parameter,
say $h$, analogous to $f$. The action possesses two extrema: at
$h=0$ and for $|h|=1$.

It turns out that the potentials (\ref{A-SU2}), (\ref{A-SU2-1}) and
(\ref{A-SU2-2}) are related to known non-abelian monopole solutions
in this theory. Namely, for $n \neq \pm 1$ and
for $n = \pm 1$ with $f=0$ these expressions coincide with the monopole
solutions with the monopole number $\kappa = n$. In fact the
solution with $\kappa = n > 0$ can be transformed
to the solution with $\kappa = -n $ by the gauge transformation
$A \rightarrow S^{-1} A S$ with the constant matrix $S=-i \tau_{1}$.
Eqs. (\ref{A-SU2}) and Eqs. (\ref{A-SU2-1}) and (\ref{A-SU2-2})
with $f=0$ describe all monopoles in the $SU(2)$ gauge theory
\cite{GNO}. As it was shown in \cite{BrNe}, \cite{Coleman} all of them,
except the trivial configuration with $n=0$, are unstable.
This is in accordance with the topological classification
of monopoles \cite{Lub} (see also \cite{Coleman}) by
elements of $\pi_{1}(G)$. This also agrees with the fact that
there is only one bundle (up to equivalence) with the base space
$S^{2}$ and the structure group $SU(2)$. The latter
statement follows from our discussion in Sect. 2. Indeed, in the
case under consideration ${\cal B}_{SU(2)}(S^{2}) \cong H^{2}(S^{2},
\pi_{1}(SU(2))) = 0$ since $\pi_{1}(SU(2))=0$.

Thus, all the monopoles are described by connections
in the trivial principal fibre bundle $P(S^{2},SU(2))$ and
can be represented by a unique form on the whole
sphere \cite{CHM}. This is indeed the case.
Namely there exist gauge transformations,
different for $U_{1}$ and $U_{2}$ patches, so that the tranformed
potentials coincide. Let us demonstrate this for the case $n=1$.
In fact this property is true for the whole family of the
invariant connections (\ref{A-SU2-1}), (\ref{A-SU2-2}).
The group elements of these gauge transformations of
the potentials on $U_{1}$ and $U_{2}$  are
\[
V_{1} = i \left( \begin{array}{cc}
 \cos \frac{\vartheta}{2} & e^{-i\varphi} \sin \frac{\vartheta}{2}  \\
 e^{i\varphi} \sin \frac{\vartheta}{2}  & - \cos \frac{\vartheta}{2}
     \end{array}  \right),
\]
\[
V_{2} = i \left( \begin{array}{cc}
 e^{i\varphi} \cos \frac{\vartheta}{2}  & \sin \frac{\vartheta}{2}  \\
 \sin \frac{\vartheta}{2}  & - e^{-i\varphi} \cos \frac{\vartheta}{2}
     \end{array}  \right).
\]
By calculating
\[
   A^{(i)'}_{1}= V_{i}^{-1} A^{(i)}_{1} V_{i} +
   \frac{1}{ie} V_{i}^{-1} d V_{i}
\]
for $i=1$ and $i=2$ one can easily check that the
transformed potentials are equal to each other and are given by
\beq
A^{(1)'}_{1}=A^{(2)'}_{1} =  \frac{1}{2e} \left( \tau_{+} c_{+} +
 \tau_{-} c_{-} + \tau_{3} c_{3} \right),   \label{A-SU2-gen}
\eeq
where
\bea
c_{+} & = & c_{-}^{*} = e^{-i\varphi} \left\{ \left[
- \cos \vartheta + \left( f \cos ^{2} \frac{\vartheta}{2}
- f^{*} \sin^{2} \frac{\vartheta}{2} \right)
 \right] \sin \vartheta d\varphi
                                         \right. \nonumber \\
 & + & \left. i(-1 + f \cos ^{2} \frac{\vartheta}{2} +
     f^{*}\sin^{2} \frac{\vartheta}{2} ) d\vartheta \right\}
                                  \label{A-SU2-gen1} \\
 c_{3} & = & \left( 1 - \frac{f + f^{*}}{2} \right)
 \sin^{2}\vartheta d\varphi -
i\frac{f-f^{*}}{2} \sin \vartheta d\vartheta.   \nonumber
\eea
Note that in general expressions (\ref{A-SU2-1}), (\ref{A-SU2-2}) and
(\ref{A-SU2-gen}) the phase of the complex parameter $f$ can be
rotated by residual gauge transformations which form the group
$U(1)$.

For $f=0$ this formula gives the known expression for the $\kappa=1$
$SU(2)$-monopole \cite{CHM}:
\beq
A^{(1)'}_{1}=A^{(2)'}_{1} = \frac{1}{4e} \left( \begin{array}{cc}
(1-\cos 2\vartheta) d\varphi & e^{-i\varphi}
(-2id\vartheta - \sin 2\vartheta d \varphi) \\
e^{i\varphi}(2id\vartheta - \sin 2\vartheta d \varphi) &
-(1-\cos 2\vartheta) d \varphi
     \end{array}  \right).  \label{A-SU2-0}
\eeq
Of course, the forms (\ref{A-SU2-gen}) and (\ref{A-SU2-0})
do not have singularities on the whole sphere.
For $f=f^{*}=1$ the forms (\ref{A-SU2-gen1}) vanish. This shows
that this configuration, which is also the extremum of the action,
describes the trivial case of the $SU(2)$-monopole with $\kappa=0$.
Note that in the original form the potentials
(\ref{A-SU2-1}) and (\ref{A-SU2-2}) do not seem to be trivial.
Of course, one can check that they are pure gauges and correspond
to the flat connection. Vanishing of the gauge field
(\ref{F-SU2}) for this value of $f$ confirms this.

The picture we obtained is the following. For different
homomorphisms $\chi_{n}: H \rightarrow G$ we constructed
different invariant connections given by Eqs. (\ref{A-SU2}),
(\ref{A-SU2-1}) and (\ref{A-SU2-2}). For $n \neq \pm 1$
or $n = \pm 1$ with $f=0$ the connection describes
the $SU(2)$-monopole solution with the monopole number $\kappa=n$.
All $SU(2)$-monopoles on $S^{2}$ are reproduced in this way.
As it was said above the solutions with numbers
$\kappa$ and $(-\kappa)$ are gauge equivalent. In addition, there is a
continuous 1-parameter family of invariant connections which passes
through the configurations describing the $SU(2)$-monopoles with
numbers $\kappa=-1$, $\kappa=0$ and $\kappa=1$
in the space of all connections of the theory. Connections
from this family are described by Eqs. (\ref{A-SU2-1}), (\ref{A-SU2-2}),
(\ref{A-SU2-gen}) and (\ref{A-SU2-gen1}). Not all of these
connections are gauge inequivalent. Classes of gauge equivalent
invariant connections are labelled by values of $|f|$. Thus, $|f|=0$
corresponds to the class of the $\kappa=1$ monopole.
The monopole with $\kappa=-1$ can be obtained from it by the gauge
transformation with the constant matrix $S= - i \tau_{1}$, as
it was explained above, and, hence, belongs to the same gauge class.
Connections with $|f|=1$ form the class describing the monopole with
$\kappa=0$.

We will study the Wilson loop functional for this
family of invariant connections shortly. But before let us consider
the third example.

\noindent \underline{Example 3. $G=SO(3)$.}
Let us recall that $SO(3) = SU(2)/Z_{2}$ and $\pi_{1}(SO(3))=Z_{2}$.
According to the discussion in Sect. 2
\[
   {\cal B}_{SO(3)}(S^{2}) \cong H^{2}(S^{2},Z_{2}) \cong Z_{2},
\]
so there are two non-equvalent principal fibre bundles
with the base $S^{2}$
and the structure group $SO(3)$: the trivial bundle
$P=S^{2} \times SO(3)$ and the non-trivial one. This is
in a accordance with the topological classification of monopole
solutions considered in Refs. \cite{Lub}, \cite{Coleman}.

For the description of $G=SO(3)$ it is convenient
to continue using $2 \times 2$ matrices with the
identification of elements $g$ and $(-g)$. In this case from
the explicit form of the group homomorphisms, Eq. (\ref{tau-hom}) it
follows that $2\kappa$ must be integer.
As in the previous example the invariant connection is given
by Eqs. (\ref{A-SU2}) for integer $n \neq \pm 1$ and
half-integer $n$ and by Eqs. (\ref{A-SU2-1}), (\ref{A-SU2-2}) and
(\ref{A-SU2-gen}) for $n = \pm 1$. For $\kappa =
n \in Z$ we have the same set of $SU(2)$-monopole solutions
with integer monopole number and the 1-parameter family of
invariant connections as before. They are connections in the
trivial bundle $P=S^{2} \times SO(3)$. For half-integer
$\kappa = n + 1/2$, $n \in Z$ the intertwining operator
$\phi = 0$. Eqs. (\ref{A-SU2}) with $n$ being substituted by $n+1/2$
describe monopoles with half-integer number $\kappa=n+1/2$.
By analyzing these potentials at the equator of the sphere
one can easily check that
they are connections in the non-trivial bundle
$P(S^{2},SO(3))$. Only the monopoles with $\kappa =0$
and $\kappa=1/2$ are stable \cite{BrNe}, \cite{Coleman}.
The former belongs to the trivial and the latter belongs
to the non-trivial topological sector of the theory, which
in their turn correspond to the trivial and non-trivial
$SO(3)$-bundles over $S^{2}$ respectively.

\vspace{0.3cm}

See \cite{invcon} for more examples and alternative descriptions
of the invariant connections.

\vspace{0.3cm}

Let us return to the example with $M=S^{2}$ and the gauge group
$G=SU(2)$ and calculate the contribution of the invariant connections
to the functional integral giving the vacuum expectation value of
the Wilson loop functional. For this we consider a family of loops
$\gamma (\vartheta_{0})$ on $S^{2}$ labelled by the angle
$\vartheta_{0}$ given by
\[
    \gamma (\vartheta_{0}) = \{(\vartheta_{0}, \varphi'),
    \vartheta_{0} = \mbox{const},
     0 \leq \varphi' < 2\pi \}.
\]
These closed paths are parallel to the equator and are parametrized
by the polar angle $\varphi'$. Here for definiteness we consider
the case when $\gamma (\vartheta_{0})$ lie in the neighbourhood
$U_{1}$, i.e. $0 \leq \vartheta_{0} \leq \pi$.

First we calculate the holonomy
\beq
  H_{\gamma (\vartheta_{0})} (A) =
  {\cal P} \exp \left[ie \int_{\gamma (\vartheta_{0})} A \right],
  \label{hol-SU2}
\eeq
where we denoted $A^{(1)}_{1}$, given by Eq. (\ref{A-SU2-1}), as $A$.
For a fixed value of $\vartheta_{0}$ we introduce the element
\[
   U(\varphi) = {\cal P}
   \exp \left[ ie \int_{\eta(\varphi;\vartheta_{0})}
   A_{\varphi} (\vartheta_{0}, \varphi') d\varphi' \right],
\]
where we indicated explicitly the dependence of the $\varphi$-component
of the gauge potential on the coordinates on the sphere.
The path $\eta(\varphi;\vartheta_{0})$ on $S^{2}$ is defined by
\[
\eta (\varphi; \vartheta_{0}) =
\{(\vartheta_{0}, \varphi'), \vartheta_{0} = \mbox{const},
     0 \leq \varphi' < \varphi \}
\]
Of course, $\eta (2\pi; \vartheta_{0}) = \gamma (\vartheta_{0})$ and
$H_{\gamma (\vartheta_{0})} (A) = U(2\pi)$.
It is easy to check that $U(\varphi)$ satisfies the matrix equation
\beq
    \frac{U(\varphi)}{d\varphi} = ie U(\varphi)  A_{\varphi}
    (\vartheta_{0},\varphi),
                 \label{U-eq}
\eeq
{}From Eq. (\ref{A-SU2-1}) we see that the dependence of $A$
on $\varphi$ is quite simple: $\varphi$ enters only
through $\exp(\pm i \varphi)$ in
the off-diagonal terms. This suggests that it can be compensated
by the unitary transformation with a matrix $T(\varphi)$:
\[
   A_{\varphi} (\vartheta_{0},\varphi) = T(\varphi)
   A_{\varphi} (\vartheta_{0},0)
   T^{-1}(\varphi).
\]
Indeed the matrix $T(\varphi)$, equal to
\begin{displaymath}
T(\varphi) = \left( \begin{array}{cc}
  e^{-i\frac{\varphi}{2}} & 0  \\
   0  & e^{i\frac{\varphi}{2}}
     \end{array}  \right),
\end{displaymath}
realizes the necessary transformation.
It is convenient to re-write Eq. (\ref{U-eq}) for the matrix
$V(\varphi):= U(\varphi) T(\varphi)$. We get
\beq
\frac{d V(\varphi)}{d \varphi} =  ie V(\varphi) A_{\varphi}
(\vartheta_{0},0)
 + V(\varphi) T^{-1}(\varphi) \frac{d T (\varphi)}{d \varphi} =
 ie V(\varphi) M(\vartheta_{0}),
               \label{V-eq}
\eeq
where the matrix $M$ does not depend on $\varphi$ and is equal to
\[
 M(\vartheta_{0}) =  A_{\varphi} (\vartheta_{0},0) - \frac{1}{2e}
 \left( \begin{array}{cc}
  1 & 0  \\
   0  & -1
     \end{array}  \right).
\]
Using Eq. (\ref{A-SU2-1}) we re-write it as
\[
          M = C_{i}(\vartheta_{0}) \tau_{i}
\]
with
\[
C_{1}(\vartheta_{0}) = - \frac{1}{4e} (f + f^{*})
   \sin \vartheta_{0}, \; \;
C_{2}(\vartheta_{0}) = - \frac{i}{4e} (f - f^{*})
   \sin \vartheta_{0}, \; \;
C_{3}(\vartheta_{0}) = - \frac{1}{2e} \cos \vartheta_{0}.
\]
The solution of Eq. (\ref{V-eq}) can now be
easily found. One obtains
\[
    V(\varphi) = e^{ie M(\vartheta_{0}) \varphi}
\]
and from this calculates the holonomy (\ref{hol-SU2}):
\beq
H_{\gamma (\vartheta_{0})} (A) = V(2 \pi) T^{-1} (2 \pi)
= - e^{2i\pi e M(\vartheta_{0}) }.   \label{hol-SU2-1}
\eeq

The traced holonomy is equal to
\bea
T_{\gamma (\vartheta_{0})} (A) & = & - \frac{1}{2}
  Tr e^{2\pi ie C_{i}(\vartheta_{0}) \tau_{i}} =
  - \cos \left( 2\pi e|C| \right) =
  - \cos \left( \pi \sqrt{\cos^{2} \vartheta_{0} +
  |f|^{2} \sin^{2} \vartheta_{0} }  \right)  \nonumber \\
  & = & - \cos \left( \pi \sqrt{ 1 + \frac{4S}{V} \left(
  1 - \frac{S}{V} \right) \left( |f|^{2} -1 \right) } \right),
  \label{T-SU2}
\eea
where $S = 2 \pi R^{2} (1 - \cos \vartheta_{0})$ is the area
of the surface surrounded by the loop $\gamma (\vartheta_{0})$ and
$V = 4 \pi R^{2}$ is the total area of the two-dimensional sphere
with the radius $R$. For $|f|=1$ $T_{\gamma (\vartheta_{0})}=1$
as it should be for the flat connection. Note that expression
(\ref{T-SU2}) is invariant under the transformation
$S \rightarrow (V-S)$.

We would like to mention that result (\ref{hol-SU2-1})
can also be obtained using an alternative technique
based on the representation of
$H_{\gamma}(A)$ in terms of a functional integral over the
Grassman variables
\cite{Grass} (see also \cite{BT2}).

Finally we calculate the quantity
\beq
  < T_{\gamma (\vartheta_{0})}>_{inv} =
  \frac{Z_{inv}( \gamma (\vartheta_{0}) )}
  {Z_{inv}(0)},    \label{T-SU2-inv}
\eeq
characterizing the contribution of invariant connections
to the vacuum expectation value of the Wilson loop functional.
The quantity $Z_{inv}( \gamma (\vartheta_{0}))$ is given
by the formula
\bea
  Z_{inv}( \gamma (\vartheta_{0}) ) & = & \int df df^{*}
  e^{-{\cal S}_{inv}(f)}
   T_{\gamma (\vartheta_{0})} (A) \nonumber \\
  & = & - \int df \  df^{*} e^{-\frac{2 \pi^{2}}{e^{2}V}
  \left( |f|^{2} - 1\right)^{2} }
  \cos \left( \pi \sqrt{ 1 + \frac{4S}{V} \left(
  1 - \frac{S}{V} \right) \left( |f|^{2} -1 \right) } \right)
  \label{Z-inv}
\eea
and is an analog of the functional integral $Z(\gamma)$,
Eq. (\ref{Zg-def}).
It mimics the ``path-integral quantization" of the
gauge model where the configuration space (the space of
connections) is finite-dimensional and consists of $SU(2)$ - invariant
connections. The integral takes into
account the contribution of the monopoles with $\kappa=0,\pm 1$ and
fluctuations around them along the {\it invariant} direction.
The action for the invariant connection in the formula
above is given by Eq. (\ref{S-SU2}).

Of course, the complete contribution of the invariant connections,
described by Eqs. (\ref{A-SU2-1}), (\ref{A-SU2-2}) or Eq.
(\ref{A-SU2-gen}), to the true functional integral (\ref{Zg-def})
is different because it includes also contributions of all
fluctuations around them. Calculation of the complete contribution is
beyond the scope of the present article. In the next section we
are going to discuss which part of the true vacuum expectation
value $<T_{\gamma (\vartheta_{0})}>$ is captured by
$<T_{\gamma (\vartheta_{0})}>_{inv}$.

We would like to make a few remarks. In Ref. \cite{Mottola} it was
argued that the Faddeev-Popov determinant on invariant
connections turns out to be zero, hence the contribution of such
connections to the functional integral vanishes. This issue was
analyzed in \cite{AGV} in a different context, namely the
authors considered similar ``quantization"
for $SU(2)$-invariant connections in $(3+1)$ - dimensional
Ashtekar's gravity. They found that zeros of the
Faddeev-Popov determinant, which are responsible for the
vanishing of the determinant, are cancelled by the contribution
of the delta functions of constraints and the path
integral measure is regular on invariant connections.

\vskip 0.1cm
\begin{figure}[ht]
\epsfxsize=0.9\hsize
\epsfbox{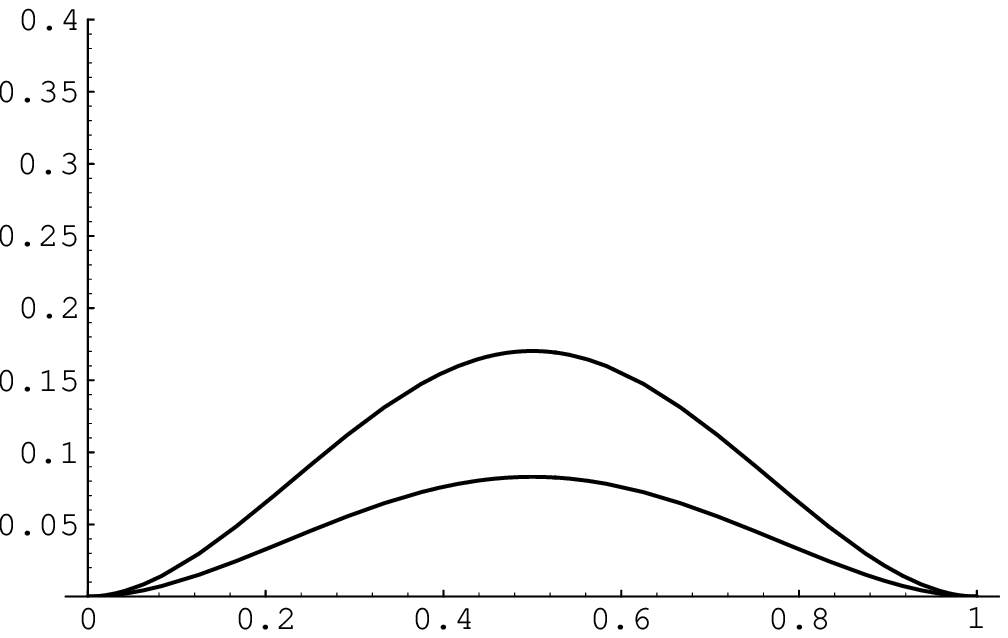}
\caption{Contribution of the invariant connections $E_{inv}(e^{2}V,S/V)$
as a function of $S/V$ for $e^{2} V=10$ (lower line)
and $e^{2} V=20$ (upper line) in the $SU(2)$ Yang-Mills theory.}
\end{figure}

The plot of the energy $E_{inv}(e^{2}V,S/V) \equiv -\ln < T_{\gamma
(\vartheta_{0})}>_{inv}$ as a function of $S/V$ for two values of
$e^{2}V$ is given in Fig. 5.
The analysis of Eqs. (\ref{T-SU2-inv}) and (\ref{Z-inv}) shows that
this function has the quadratic behaviour in $S/V$ for
$0 < S/V < (S/V)_{*} < 1/2$ (the value of $(S/V)_{*}$ depends on
$(e^{2}V)$ and is about $0.35$ for the plots in Fig. 5).
Recall that we found similar behaviour for the contribution
of the abelian monopoles in Sect. 3.

\section{Discussion of the results}

In the present article we studied the vacuum expectation value
$<T_{\gamma}>$ of the Wilson loop variable in two-dimensional
pure Yang-Mills theories. Firstly we discussed the calculation
of this quantity by various techniques.
In the abelian case we expanded the functional
integral giving $<T_{\gamma}>$ in a sum over topological
sectors of the theory. They are in 1-1 correspondence with
non-equivalent principal fibre bundles $P(M,G)$. As a by-product we
obtained a classification of such bundles in terms of elements
of the cohomology group $H^{2}(M,\pi_{1}(G))$, Eq. (\ref{BG}).
In fact, this classification is valid even for the case when
$M$ is a two-dimensional CW-complex. The integrals over
connections in each topological sector were calculated semiclassically
by expanding the action around the monopole configuration,
representing this sector, and performing the Gaussian integration
over the fluctuations. Of course, since the theory is abelian,
the semiclassical approximation is exact. This calculation was
done entirely within the geometrical description of gauge theories
in terms of gauge connections in principal fibre bundles and
for an arbitrary compact two-dimensional space-time $M$. This
approach also allowed us to separate the contribution of the
monopoles.

In the case of an arbitrary gauge group $G$ the calculation
was carried on the lattice. We performed it directly in terms of
plaquette variables (holonomies for loops associated to the
plaquettes). Relation (\ref{homot}) between the loops for a given
two-dimensional manifold induces the corresponding relation between
the plaquette variables and the holonomies for homotopically
non-trivial loops which was imposed by inserting the
$\delta$-function. As far
as we know, such technique was not used before for the
calculation of $<T_{\gamma}>$. Of course, our result (\ref{Tlat})
agrees with the results of previous calculations \cite{W-calc} -
\cite{ALMMT}. General expression for $<T_{\gamma}>$, Eq. (\ref{Tlat}),
being specified to the gauge
group $G=U(1)$, reproduces the result of Sect. 3 thus proving that
the functional integral for the abelian case is indeed saturated
by the sum over the topological sectors and does not contain any
contributions different from the instanton
contributions and fluctuations around them (including, of course,
the trivial perturbative sector).

In accordance with results of previous studies, we confirm that
$<T_{\gamma}>$ has the area law behaviour which is a characteristic
feature of two-dimensional pure Yang-Mills theories. Namely,
$E(e^{2}V, S/V) = -\ln <T_{\gamma}>$ is linear in
$S/V$, where $S$ is a region surrounded by the loop $\gamma$, $V$
is the total area of the surface and $e$ is the gauge coupling,
in the interval $0 < S/V < (S/V)_{*}$ (remember that $M$ is a
compact surface). The range where the function $E$ is almost linear
depends on the value of $e^{2}V$, and the larger $e^{2}V$, the closer
to $1/2$ $(S/V)_{*}$ is. Thus for $e^{2}V=20$ $(S/V)_{*}
\approx 0.45$. The question we analyzed in this article is
the contribution of the monopoles and the invariant connections
to $E(e^{2}V, S/V)$.

The example with $G=U(1)$ shows that the contribution of the monoples
depends quadratically on $S/V$ for $0 < S/V < (S/V)_{*}$. The
linear behaviour of $E(e^{2}V, S/V)$ for $(S/V) \ll 1$ comes
from the contribution of fluctuations. However, for $S/V \sim (S/V)_{*}$
higher corrections in $S/V$ already become important and lead
to the deviation from the area law behaviour. The restauration
of the area law for the complete function $E(e^{2}V, S/V)$ is the
result of exact cancellation of the quadratic terms in $S/V$ in the
sum of the contributions of the monoples and fluctuations. In this
sense the quadratic dependence of the contribution of monopoles
$E_{mon}(e^{2}V, S/V)$ is an indicator of the area law behaviour
of the complete function.

Similar analysis can be carried out in the non-abelian case.
For $G=SU(2)$ and $G=SO(3)$ the plots of $E(e^{2}V, S/V)$ and
$E_{mon}(e^{2}V, S/V)$ are qualitatively the same as in Fig. 1
and we do not present them in this article. The only difference
is the interval of the values of the functions. For
$E_{mon}(e^{2}V, S/V)$ we took just the sum
$\sum_{n} \exp \left( - {\cal S}(\tilde{A}^{(n)}) \right)$
of the leading quasiclassical contributions of the non-abelian
monopole solutions found in Sect. 5, where ${\cal S}(\tilde{A}^{(n)})$
is the action on these monopole solutions. We did not carry out
the detailed analysis of contributions of fluctuations
around the monopoles in these cases. Apparently the similarity with
the abelian case arises from the fact that in a special gauge (say,
$A_{1} = 0$) the action of the theory becomes quadratic and
we can expect the factorization of the contribution of fluctuations
as in Eq. (\ref{Z-fact}). We argue that again the quadratic behaviour
of $E_{mon}(e^{2}V, S/V)$ in $S/V$ for $0 < S/V < (S/V)_{*}$ serves
as an indicator of the power law behaviour.

Finally, we studied the class of invariant connections. In the case
when the space-time is the two-dimensional sphere, $M = S^{2}$,
and for the gauge groups $G=U(1)$, $SU(2)$ and $SO(3)$ we
constructed all such connections and showed that they include
all monopole solutions existing in these theories. We saw that for
the non-abelian cases in addition to this the class of invariant
connections contains a continuous family parametrized by a
finite number of parameters. Remembering that invariant
connections play important role for certain problems of field
theory, we addressed the question which features of the vacuum
expectation value $<T_{\gamma}>$ of the Wilson loop variable
are captured by the sector of such connections. For this we
calculated $<T_{\gamma}>$ for the continuous family of invariant
connections. Since the space of connections
is finite dimensional the functional integral reduces to an
ordinary one. The function $E_{inv}(e^{2}V, S/V)$ for such
"gauge theory" with finite number of degrees of freedom
again has the quadratic behaviour in $S/V$.

It is interesting to study whether such relation between the
quadratic behaviour of the contribution of invariant connections
and the area law dependence of the complete function
$E(e^{2}V, S/V)$ maintains in higher dimensions. If this is the case,
then an indication of the presence or absence of the area law
dependence of the complete theory can be obtained from the analysis
of the much simpler finite dimensional sector of invariant
connections. We plan to study this issue in a future publication.

\bigskip

\noindent{\large \bf Acknowledgments}

We would like to thank M. Asorey, A. Demichev, J. Mour\~ao,
R. Picken and S. Xamb\'o
for useful discussions and valuable remarks. 
The authors acknowledge financial support from DGESIC, M.E.C., Spain 
(project HP98-0040) and from Conselho de Reitores das Universidades
Portuguesas (Ac{\c c}\~oes Integradas Luso-Espanholas, E-79/99). 
Yu.K. acknowledges financial support from Russian Fund for 
Basic Research (grant 98-02-16769-a) and fellowship 
PRAXIS XXI/BCC/4802/95.

\bigskip

\section*{Appendix A. Differential calculus on the lattice and
calculation of the Wilson loop}
\def\theequation{A\arabic{equation}}
\setcounter{equation}{0}

In this section we present an alternative technique for
the calculation of $<T_{\gamma}>$ in terms of differential
forms on the lattice \cite{gu}, \cite{abf} which is applicable
for abelian gauge groups. Here we consider the case $G=U(1)$.

Lattice $q$-forms are functions defined on the $q$-cells
of a two-dimensional lattice $\Lambda$ with values in
an abelian group $K$. The set of such forms is
denoted by $\Lambda^{q}(\Lambda , K)$, $q=0,1,2$.

The loop $\gamma$ is characterized by the integer valued 1-form
$J_\gamma$ whose values $J_\gamma(l)$ give the number of times
the loop passes through the link $l \in \Lambda_{1}$
(taking into account the orientation).
The gauge potential is charcterized by the 1-form $\theta$ whose
values are real numbers modulo $2\pi$. It is related to the
gauge potential and the variables introduced in Sect. 4 through
\beq
g_l = e^{i\theta (l)} = e^{ie \int_{l} A}, \; \; \;
g_{\gamma_p} =\prod_{l\in p} g_l, \label{g-pl}
\eeq
where $l$ and $p$ denote links and plaquettes of $\Lambda$ respectively.

The operators of the continuous case have their discrete counterparts
on the lattice. Let us introduce the exterior derivative operator $d$
and its adjoint $\delta$ which map $q$-forms to $q+1$ and $q-1$
forms respectively:
\[
d : \Lambda^{q}(\Lambda,K) \longrightarrow \Lambda^{q+1}(\Lambda,K),
\; \; \;
\delta : \Lambda^{q}(\Lambda,K) \longrightarrow \Lambda^{q-1}(\Lambda,K).
\]
The action of $d$ and $\delta$ is defined by:
\bea
df(c_{q+1}) & = & \sum_{c_q}I(c_{q+1},c_q)f(c_q),   \nonumber \\
\delta f(c_{q-1}) & = & \sum_{c_q}I(c_q,c_{q+1})f(c_q), \nonumber
\eea
where $f \in \Lambda^{q}(\Lambda, K)$, $c_{q} \in \Lambda_{q}$ and
$I$ is the incidence function introduced in Sect. 4.
The operator $d$ is zero on $2$-forms and $\delta$ is zero on $0$-forms.
Let us introduce now the scalar product of two $q$-forms $f$ and $h$ as
$<f,h> = \sum_{c_{q}} f(c_{q}) h(c_{q})$.
It is easy to verify that $\delta$ and $d$ are adjoint to
each other with respect to the scalar product:
$\langle f,dh\rangle=\langle \delta f,h\rangle$,
where $f$ is a form one degree higher than $h$.
The Laplacian operator $\Delta$, mapping $q$-forms into
$q$-forms, is defined in a standard way as
$\Delta =\delta d + d\delta$. It is self-adjoint,
$\langle f,\Delta g\rangle = \langle \Delta f,g\rangle $,
and positive-semidefinite:
\[
\langle f,\Delta f\rangle =\langle df,df\rangle +
\langle \delta f,\delta f\rangle =
\| d f\|^2+\| \delta f\|^2.
\]
One can easily check that the plaquette variable $g_{\gamma_p}$ in Eq.
(\ref{g-pl}) is equal to
\[
g_{\gamma_p} =e^{id\theta (p)}.
\]

It can be shown that alternatively to the loop variables
and the measure used in Sect. 4, one can take $\theta$ as the
integration variables and $D\theta \equiv \prod_l d (\theta (l))$
as the measure. Then the path integral (\ref{Zp-def}) can be written as
\[
Z_{\Lambda}(\gamma ) = \int D\theta \prod_p \exp
\{-{\cal S}_{p}(g_{\gamma_{p}}) \}
\exp (i<J_\gamma,\theta >).
\]
The action can be expressed through the Fourier coefficients $B_{n}(p)$
as in Eq. (\ref{L-action1}). Remember that now we are considering the case
$G=U(1)$, so the irreducible representations are labelled by integers.
Since the action is symmetric $B_{-n}(p)=B_n(p)$). Then we have
\bea
Z_{\Lambda}(\gamma ) & = & \int D\theta \prod_{ p }
\sum_{n_p}B_{n_p}(p)\exp \left( -in_{p}d\theta (p) \right)
\exp \left( i<J_{\gamma},\theta > \right)  \nonumber \\
 & = & \int D\theta \sum_{\{n_p\}}\prod_{ p } [B_{n_p}(p)
\exp \left( -in_{p}d\theta (p) \right)]
\exp \left( i<J_{\gamma},\theta > \right)   \nonumber \\
 & = & \int D\theta \sum_{s}(\prod_{ p } B_{s(p)}(p))
\exp \left( -i<s,d\theta >+i<J_{\gamma},\theta > \right) \nonumber  \\
 & = & \int D\theta \sum_{s}(\prod_{ p } B_{s(p)}(p))
\exp \left( -i<\delta s-J_{\gamma},\theta > \right),   \nonumber
\eea
where we have written  the sum over configurations $\{n_p\}$
(a configuration is a set of integers $n_{p}$ associated to the
plaquettes) as a sum oves integer valued 2-forms
$s \in \Lambda^{2}(\Lambda,Z)$. These 2-forms characterize
surfaces on $M$. By integrating over $\theta (l)$ we obtain that
\[
Z_{\Lambda}(\gamma ) = \sum_{s}(\prod_{ p } B_{s (p)}(p))
\delta (\delta s-J_\gamma )
= \sum_{s: \: \delta s=J_\gamma}(\prod_{ p } B_{s(p)}(p)).
\]
In the sum above only the 2-forms which correspond to the surfaces
whose border is the loop $\gamma$ contribute.
The expectation value of the Wilson loop finally reads:
\[
<T_\gamma >=\frac{\sum_{s: \: \delta s=J_\gamma}(\prod_{ p }
B_{s(p)}(p))} {\sum_{s: \: \delta s=0}(\prod_{ p } B_{s(p)}(p))}.
\]

For the Villain action (see Eq. (\ref{hk})) we obtain
\[
<T_\gamma >=\frac{\sum_{s: \: \delta s=J_{\gamma}}
\exp \left\{-\frac{e^2}{2}\sum_p (s(p))^{2}|p| \right\}}
{\sum_{s: \: \delta s=0}\exp \left\{-\frac{e^2}{2}\sum_p
(s(p))^{2}|p| \right\}}.
\]
In the compact case due to the condition $\delta s=0$ only the
surfaces which wrap the whole lattice $n$ times ($n\in Z$),
i.e. whose 2-forms satisfy $s(p)=n$ for all
$p \in \Lambda_{2}$, contribute to the denominator.
If the loop $\gamma$ is simple and is the boundary of a region
in $M$, then the condition $\delta s=J_{\gamma}$ implies that
the only forms which give non-zero contribution are those which
have the following two properties:
1) $s$ is constant inside the regions bordered by $\gamma$,
i.e. $s(p)=s(p')$ if both $p$ and $p'$ lie either in the interior
or in the exterior of $\gamma$;
2) $s(p) - s(p') = \pm 1$ (the sign
depends on the relative orientation) when one of the plaquettes
$p$ and $p'$ lies in the interior of the loop $\gamma$ and the
other lies in the exterior of it.
With this results (\ref{Tg}) and (\ref{Tlat}) are recovered.

\end{document}